\documentclass[twocolumn]{aastex631}
\usepackage[utf8]{inputenc}
\usepackage{natbib}
\usepackage{graphicx}
\usepackage{xcolor}
\usepackage{threeparttable}

\usepackage{newtxtext,newtxmath}
\usepackage{amsmath}	
\usepackage{multirow}
\usepackage{tabularx}

\received{2024 September 26}
\revised{2025 February 28}
\accepted{2025 March 19}
\submitjournal{AJ}

\shorttitle{WASP-80\,{\rm b} NIRISS/SOSS Emission}
\shortauthors{Morel et al.}

\begin{document}

\title{A Moderate Albedo from Reflecting Aerosols on the Dayside of WASP-80\,b Revealed by JWST/NIRISS Eclipse Spectroscopy}


\correspondingauthor{Kim Morel}
\email{kim.morel@umontreal.ca}


\author[0000-0002-1901-1266]{Kim Morel}
\affiliation{Institut Trottier de recherche sur les exoplanètes, 1375 Avenue Thérèse-Lavoie-Roux, Montréal, QC, H2V 0B3, Canada}
\affiliation{Département de Physique, Université de Montréal, 1375 Avenue Thérèse-Lavoie-Roux, Montréal, QC, H2V 0B3, Canada}
\affiliation{Department of Physics, McGill University, 3600 rue University, Montréal, QC, H3A 2T8, Canada}

\author[0000-0002-2195-735X]{Louis-Philippe Coulombe}
\affiliation{Institut Trottier de recherche sur les exoplanètes, 1375 Avenue Thérèse-Lavoie-Roux, Montréal, QC, H2V 0B3, Canada}
\affiliation{Département de Physique, Université de Montréal, 1375 Avenue Thérèse-Lavoie-Roux, Montréal, QC, H2V 0B3, Canada}

\author[0000-0002-5904-1865]{Jason F. Rowe}
\affiliation{Institut Trottier de recherche sur les exoplanètes, 1375 Avenue Thérèse-Lavoie-Roux, Montréal, QC, H2V 0B3, Canada}
\affiliation{Department of Physics \& Astronomy, Bishop’s University, Sherbrooke, QC, J1M 1Z7, Canada}

\author[0000-0002-6780-4252]{David Lafrenière}
\affiliation{Institut Trottier de recherche sur les exoplanètes, 1375 Avenue Thérèse-Lavoie-Roux, Montréal, QC, H2V 0B3, Canada}
\affiliation{Département de Physique, Université de Montréal, 1375 Avenue Thérèse-Lavoie-Roux, Montréal, QC, H2V 0B3, Canada}

\author[0000-0003-0475-9375]{Loïc Albert}
\affiliation{Institut Trottier de recherche sur les exoplanètes, 1375 Avenue Thérèse-Lavoie-Roux, Montréal, QC, H2V 0B3, Canada}
\affiliation{Département de Physique, Université de Montréal, 1375 Avenue Thérèse-Lavoie-Roux, Montréal, QC, H2V 0B3, Canada}

\author[0000-0003-3506-5667]{\'Etienne Artigau}
\affiliation{Institut Trottier de recherche sur les exoplanètes, 1375 Avenue Thérèse-Lavoie-Roux, Montréal, QC, H2V 0B3, Canada}
\affiliation{Département de Physique, Université de Montréal, 1375 Avenue Thérèse-Lavoie-Roux, Montréal, QC, H2V 0B3, Canada}

\author[0000-0001-6129-5699]{Nicolas B. Cowan}
\affiliation{Institut Trottier de recherche sur les exoplanètes, 1375 Avenue Thérèse-Lavoie-Roux, Montréal, QC, H2V 0B3, Canada}
\affiliation{Department of Physics, McGill University, 3600 rue University, Montréal, QC, H3A 2T8, Canada}
\affiliation{Department of Earth and Planetary Sciences, McGill University, 3600 rue University, Montréal, QC, H3A 2T8, Canada}

\author[0000-0003-4987-6591]{Lisa Dang}
\altaffiliation{Banting Postdoctoral Fellow}
\affiliation{Institut Trottier de recherche sur les exoplanètes, 1375 Avenue Thérèse-Lavoie-Roux, Montréal, QC, H2V 0B3, Canada}
\affiliation{Département de Physique, Université de Montréal, 1375 Avenue Thérèse-Lavoie-Roux, Montréal, QC, H2V 0B3, Canada}

\author[0000-0002-3328-1203]{Michael Radica}
\affiliation{Institut Trottier de recherche sur les exoplanètes, 1375 Avenue Thérèse-Lavoie-Roux, Montréal, QC, H2V 0B3, Canada}
\affiliation{Département de Physique, Université de Montréal, 1375 Avenue Thérèse-Lavoie-Roux, Montréal, QC, H2V 0B3, Canada}

\author[0000-0003-4844-9838]{Jake Taylor}
\affiliation{Department of Physics, University of Oxford, Parks Rd, Oxford, OX1 3PU, UK}

\author[0000-0002-2875-917X]{Caroline Piaulet-Ghorayeb}
\affiliation{Institut Trottier de recherche sur les exoplanètes, 1375 Avenue Thérèse-Lavoie-Roux, Montréal, QC, H2V 0B3, Canada}
\affiliation{Département de Physique, Université de Montréal, 1375 Avenue Thérèse-Lavoie-Roux, Montréal, QC, H2V 0B3, Canada}

\author[0000-0001-6809-3520]{Pierre-Alexis Roy}
\affiliation{Institut Trottier de recherche sur les exoplanètes, 1375 Avenue Thérèse-Lavoie-Roux, Montréal, QC, H2V 0B3, Canada}
\affiliation{Département de Physique, Université de Montréal, 1375 Avenue Thérèse-Lavoie-Roux, Montréal, QC, H2V 0B3, Canada}

\author[0000-0001-5578-1498]{Björn Benneke}
\affiliation{Institut Trottier de recherche sur les exoplanètes, 1375 Avenue Thérèse-Lavoie-Roux, Montréal, QC, H2V 0B3, Canada}
\affiliation{Département de Physique, Université de Montréal, 1375 Avenue Thérèse-Lavoie-Roux, Montréal, QC, H2V 0B3, Canada}

\author[0000-0002-7786-0661]{Antoine Darveau-Bernier}
\affiliation{Institut Trottier de recherche sur les exoplanètes, 1375 Avenue Thérèse-Lavoie-Roux, Montréal, QC, H2V 0B3, Canada}
\affiliation{Département de Physique, Université de Montréal, 1375 Avenue Thérèse-Lavoie-Roux, Montréal, QC, H2V 0B3, Canada}

\author[0000-0002-8573-805X]{Stefan Pelletier}
\affiliation{Observatoire astronomique de l'Université de Genève, 51 chemin Pegasi 1290 Versoix, Switzerland}
\affiliation{Institut Trottier de recherche sur les exoplanètes, 1375 Avenue Thérèse-Lavoie-Roux, Montréal, QC, H2V 0B3, Canada}

\author[0000-0001-5485-4675]{René Doyon}
\affiliation{Institut Trottier de recherche sur les exoplanètes, 1375 Avenue Thérèse-Lavoie-Roux, Montréal, QC, H2V 0B3, Canada}
\affiliation{Département de Physique, Université de Montréal, 1375 Avenue Thérèse-Lavoie-Roux, Montréal, QC, H2V 0B3, Canada}

\author[0000-0002-6773-459X]{Doug Johnstone}
\affiliation{NRC Herzberg Astronomy and Astrophysics, 5071 West Saanich Rd, Victoria, BC, V9E 2E7, Canada}
\affiliation{Department of Physics and Astronomy, University of Victoria, Victoria, BC, V8P 5C2, Canada}

\author[0000-0002-4451-1705]{Adam B. Langeveld}
\affiliation{Department of Physics and Astronomy, Johns Hopkins University, 3400 N. Charles Street, Baltimore, MD 21218, USA}
\affiliation{Department of Astronomy and Carl Sagan Institute, Cornell University, Ithaca, NY 14853, USA}

\author[0000-0002-1199-9759]{Romain Allart}
\affiliation{Institut Trottier de recherche sur les exoplanètes, 1375 Avenue Thérèse-Lavoie-Roux, Montréal, QC, H2V 0B3, Canada}
\affiliation{Département de Physique, Université de Montréal, 1375 Avenue Thérèse-Lavoie-Roux, Montréal, QC, H2V 0B3, Canada}

\author[0000-0001-6362-0571]{Laura Flagg}
\affiliation{Department of Physics and Astronomy, Johns Hopkins University, 3400 N. Charles Street, Baltimore, MD 21218, USA}

\author[0000-0001-7836-1787]{Jake D. Turner}
\affiliation{Department of Astronomy and Carl Sagan Institute, Cornell University, Ithaca, NY 14853, USA}

\begin{abstract}

Secondary eclipse observations of exoplanets at near-infrared wavelengths enable the detection of thermal emission and reflected stellar light, providing insights into the thermal structure and aerosol composition of their atmospheres. These properties are intertwined, as aerosols influence the energy budget of the planet. WASP-80\,b is a warm gas giant with an equilibrium temperature of 825\,K orbiting a bright late-K/early-M dwarf, and for which the presence of aerosols in its atmosphere have been suggested from previous HST and Spitzer observations. We present an eclipse spectrum of WASP-80\,b obtained with JWST NIRISS/SOSS, spanning 0.68 to 2.83\,$\mu$m, which includes the first eclipse measurements below 1.1\,$\mu$m for this exoplanet, extending our ability to probe light reflected by its atmosphere.
When a reflected light geometric albedo is included in the atmospheric retrieval, our eclipse spectrum is best explained by a reflected light contribution of $\sim$30\,ppm at short wavelengths, although further observations are needed to statistically confirm this preference.
We measure a dayside brightness temperature of $T_{\rm B}=811_{-70}^{+69}$\,K and constrain the reflected light geometric albedo across the SOSS wavelength range to $A_{\rm g}=0.204_{-0.056}^{+0.051}$, allowing us to estimate a 1-$\sigma$ range for the Bond albedo of $0.148\lesssim A_{\rm B}\lesssim0.383$. By comparing our spectrum with aerosol models, we find that manganese sulfide and silicate clouds are disfavored, while cloud species with weak-to-moderate near-infrared reflectance, along with soots or low formation-rate tholin hazes, are consistent with our eclipse spectrum.

\end{abstract}

\keywords{Exoplanets (498); Hot Jupiters (753); Exoplanet atmospheres (487); Exoplanet atmospheric composition (2021); Extrasolar gaseous giant planets (509)}

\vspace{0.25cm}


\section{Introduction}
\label{sec:Intro}

Transiting close-in ($P\leq10$\,days) gas giants are ideal targets for atmospheric characterization as their elevated temperatures and high planet-to-star radius ratios produce strong transit and eclipse spectroscopy signals. With secondary eclipse observations, it is possible to probe their thermal emission and reflected light as they pass behind their host star. Close-in giants on circular orbits are assumed to be tidally locked \citep{kasting_habitable_1993, barnes_tidal_2017}, implying that we observe their permanent dayside--the hemisphere that constantly faces the star--during secondary eclipse.

The Spitzer Space Telescope provided measurements that led to the first detection of thermal emission from a close-in exoplanet \citep{charbonneau_detection_2005, deming_infrared_2005} and has since had an enormous impact by measuring many more secondary eclipses (e.g. \citealt{line_systematic_2014, kammer_spitzer_2015, changeat_five_2022, deming_emergent_2023, dang_comprehensive_2024}), providing further insights into the thermal structure of exoplanets. More recently, the first emission spectra from data collected by the James Webb Space Telescope (JWST) were obtained \citep{coulombe_broadband_2023,bell_methane_2023, august_confirmation_2023, schlawin_multiple_2024}, enabling more precise dayside atmospheric characterization than ever before, by providing eclipse measurements over a larger wavelength range. Moreover, recent NIRISS/SOSS observations of the ultra-hot Neptune LTT\,9779\,b have demonstrated the instrument's capability to simultaneously capture thermal emission and reflected light \citep{Coulombe2025highlyreflectivewhiteclouds}.

Reflected stellar light can only be probed at visible and near-infrared (NIR) wavelengths, as most stars emit predominantly short-wavelength light and aerosols reflect more in that regime. Studying reflected light measurements can provide insights into the presence and composition of aerosols in exoplanet atmospheres. Close-in gas giants receive large ultraviolet (UV) fluxes from their host star, leading to photochemical processes that potentially create hazes, which absorb and/or reflect light \citep{line_thermochemical_2011, zahnle_thermometric_2009}. Clouds, which are formed by condensation through thermochemical equilibrium, can also be expected in the atmospheres of warm Jupiters ($500<T_{\rm eq}<1000$\,K), as many cloud species condense at higher temperatures than those of their atmospheres \citep{gao_aerosols_2021}. These clouds can be made, for example, of silicates, salts, and sulfides; species that are found in the atmospheres of brown dwarfs in this temperature regime \citep{marley_clouds_2002, morley_neglected_2012, gao_aerosols_2021}.

Aerosols have an impact on the thermal structure of an atmosphere. Their altitude and opacity determine how much stellar radiation can penetrate the atmosphere, and in turn influences its temperature-pressure (T-P) profile \citep{seager_photometric_2000, seager_exoplanet_2010-1, barstow_clouds_2014, morley_thermal_2015, keating_revisiting_2017}. Different aerosols also shape the observed spectra in distinguishable ways. In general, thick clouds reflect light with weak spectral dependence at NIR and visible wavelengths \citep{marley_reflected_1999, morley_thermal_2015}. However, if the atmosphere contains hazes, the planet can appear brighter in the NIR and darker at short wavelengths as most types of hazes are more likely to absorb blue and UV light \citep{marley_reflected_1999, morley_thermal_2015}.

To quantify these processes, the albedo of the planet must be measured. Geometric albedo spectra can be used to identify aerosol species \citep{marley_reflected_1999, sudarsky_albedo_2000}, while the Bond albedo, which can be constrained from thermal emission measurements, dictates the energy budget of the planet \citep{cowan_statistics_2011, schwartz_balancing_2015}. Warm Jupiters are expected to have low to moderate Bond albedos ($A_\mathrm{B}\lesssim0.4$, \citealt{marley_reflected_1999, cowan_statistics_2011, schwartz_balancing_2015, bell_comprehensive_2021}). In contrast, cold giants ($T_\mathrm{eq}\lesssim250$\,K) can achieve higher Bond albedos if their atmospheres develop reflective ammonia or water clouds \citep{marley_reflected_1999, sudarsky_albedo_2000, li_less_2018, macdonald_exploring_2018}.

WASP-80 b is a warm Jupiter ($T_{\rm eq}=825\pm19$\,K, $R_{\rm p}=0.999^{+0.030}_{-0.031}$\,R$_{\rm J}$, $M_{\rm p}=0.538^{+0.035}_{-0.036}$\,M$_{\rm J}$, \citealt{triaud_wasp-80b_2015}) with an orbital period of 3.07\,days \citep{triaud_wasp-80b_2015, bell_methane_2023}. It is part of the small known population of gas giants orbiting low-mass stars: WASP-80 is a bright, late-K/early-M dwarf ($R_*=0.586$\,R$_\odot$, $M_*=0.577$\,M$_\odot$) with an effective temperature of 4143\,K \citep{triaud_wasp-80b_2015}. \cite{bryant_occurrence_2023} estimated the occurrence rate of transiting giants ($0.6\leq R_{\rm p} \leq 2.0\,{\rm R_J}$) with short periods ($P\leq10$\,days) around stars with masses between 0.088 and 0.71\,M$_\odot$ to be $0.194 \pm 0.072\%$. For comparison, \cite{beleznay_exploring_2022} estimated the occurrence rate of hot Jupiters ($9<R_{\rm p}<28\,$R$_\oplus$, $0.8<P<10$\,days) around AFG dwarf stars ($0.8<M_*<2.3$\,M$_\odot$) to be $0.33\pm 0.04\%$.
This makes WASP-80\,b a uniquely important target for atmospheric study. Its relatively low equilibrium temperature also makes its reflected light spectrum well isolated from its thermal emission spectrum.

\begin{table}
\begin{center}
\caption{Parameters of the WASP-80 planetary system used in this analysis.}
\label{tab:1} 
\begin{tabular}{lcc}\hline
\textbf{Parameters} & \textbf{WASP-80} & \textbf{Ref.}\\ \hline
Spectral type & K7-M0 V & 2\\
$R_*$ (R$_\odot$) & 0.586 $^{+0.017}_{-0.018}$ & 1\\
$M_*$ (M$_\odot$) & 0.577 $^{+0.051}_{-0.054}$ & 1\\
{\rm [Fe/H]} & -0.13 $^{+0.15}_{-0.17}$ & 1\\
log \textit{g} (log$_{10}$ cm/s$^2$) & 4.663 $^{+0.015}_{-0.016}$ & 1\\
$T_{\rm eff}$ (K) & 4143 $^{+92}_{-94}$ & 1\\ \hline

\textbf{Parameters} & \textbf{WASP-80 b} & \textbf{Ref.}\\ \hline
R$_{\rm p}$ (R$_{\rm J}$) & 0.999 $^{+0.030}_{-0.031}$ & 1\\
M$_{\rm p}$ (M$_{\rm J}$) & 0.538 $^{+0.035}_{-0.036}$ & 1\\
$P$ (days) & $3.067851945\pm0.000000026$ & 3\\
$e$ & 0 & 1, 3\\
$i$ (deg) & $88.938\pm0.059$ & 3\\
$a/R_*$ & $12.643\pm0.032$ & 3\\
$R_{\rm p}/R_*$ & $0.17137^{+0.00037}_{-0.00039}$ & 1\\
$T_{\rm eq}$ (K) & $825\pm19$ & 1\\ \hline
\multicolumn{3}{l}{\footnotesize{[1] \cite{triaud_wasp-80b_2015}, [2] \cite{triaud_wasp-80b_2013}, [3] \cite{bell_methane_2023}}} \\
\end{tabular}
\end{center}
\end{table}

Using transit observations from the Hubble Space Telescope (HST) and Spitzer, \cite{wong_hubble_2022} found evidence of Rayleigh scattering produced by fine-particle hazes, as well as the presence of a cloud deck extending up to 1\,mbar on the limbs of WASP-80\,b. Using Spitzer/IRAC (3.6 \& 4.5\,$\mu$m) eclipse observations, they also concluded that the planet has a low Bond albedo. \cite{jacobs_probing_2023} used 5 secondary eclipse observations of WASP-80\,b obtained with HST WFC3/G141 ($1.1-1.7$\,$\mu$m) and found an average eclipse depth of $34\pm10$\,ppm in the WFC3/G141 bandpass. From this, they obtained a 3-$\sigma$ upper limit on the geometric albedo of $A_{\rm g}<0.33$. Their results also show a slight preference for deep dayside clouds over hazes. However, they were not able to fit a consistent model to both their eclipse spectrum and the previously published HST/Spitzer transmission spectrum, indicating heterogeneous aerosol distribution between the dayside and limbs. \cite{bell_methane_2023} used transit and secondary eclipse observations of WASP-80\,b obtained with the F322W2 grism of the Near Infrared Camera (NIRCam) of JWST to detect strong absorption features of water and methane, in agreement with theoretical predictions on the dominance of CH$_{4}$ over CO for planets with $T_{\rm eq}\lesssim1000$\,K \citep{lodders_atmospheric_2002}. They also inferred the presence of optically thick clouds at pressures $\lesssim0.2$\,mbar (2-$\sigma$ upper limit) on the planet terminator.

In this work, we present a red-optical and NIR (0.68 - 2.83\,$\mu$m) secondary eclipse spectrum of WASP-80\,b obtained with the Near InfraRed Imager and Slitless Spectrograph (NIRISS, \citealt{doyon_near_2023}) on board JWST in its Single Object Slitless Spectroscopy (SOSS, \citealt{albert_near_2023}) observational mode. This dataset includes the first eclipse measurements of WASP-80\,b below 1.1\,$\mu$m, which allows us to put tighter constraints on its albedo and suggest the presence of aerosols in its atmosphere. We describe the observations and the data reduction in Section \ref{sec:Obs_and_reduc}. The light curve fitting is detailed in Section \ref{sec:Fitting}, and the atmosphere and cloud modeling is presented in Section \ref{sec:modeling}. We then discuss our results in Section \ref{sec:results_discussion} and conclude in Section \ref{sec:Conclusion}.\\

\section{Observations and Data Reduction}
\label{sec:Obs_and_reduc}

\subsection{Observations}
\label{sec:Observation}

A secondary eclipse of WASP-80\,b was observed on October 26, 2023, with the SOSS mode of the NIRISS instrument on board JWST as part of the NEAT (NIRISS Exploration of the Atmospheric diversity of Transiting exoplanets) cycle 1 Guaranteed Time Observations program (PID: 1201; PI: David Lafrenière). Using the GR700XD grism combined with the CLEAR filter, the time series observations (TSO) consisted of 668 integrations of 21.976 s each with four groups per integration, and started at 08:58:15.221 UTC for a total exposure duration of 5.10\,hours. The SUBSTRIP256 subarray was used to capture diffraction orders 1 and 2, covering wavelengths from 0.6 to 2.83\,$\mu$m. 

\subsection{Data reduction}
\label{sec:Reduction}

We performed a Stage 1 \& 2 reduction on the TSO using the \texttt{supreme-SPOON v1.3.0} pipeline \citep{radica_awesome_2023, feinstein_early_2023, coulombe_broadband_2023, lim_atmospheric_2023, fournier-tondreau_near-infrared_2024, radica_muted_2024, Radica2024}\footnote{Note that this pipeline is now known as \texttt{exoTEDRF} (\url{https://github.com/radicamc/exoTEDRF})} and started from the raw, uncalibrated files downloaded from the Mikulski Archive for Space Telescopes (MAST)\footnote{All the NIRISS/SOSS data used in this paper can be found in MAST: \dataset[https://doi.org/10.17909/htd7-fg43]{https://doi.org/10.17909/htd7-fg43})}.

\begin{figure*}
	\centering
	\includegraphics[width=\textwidth]{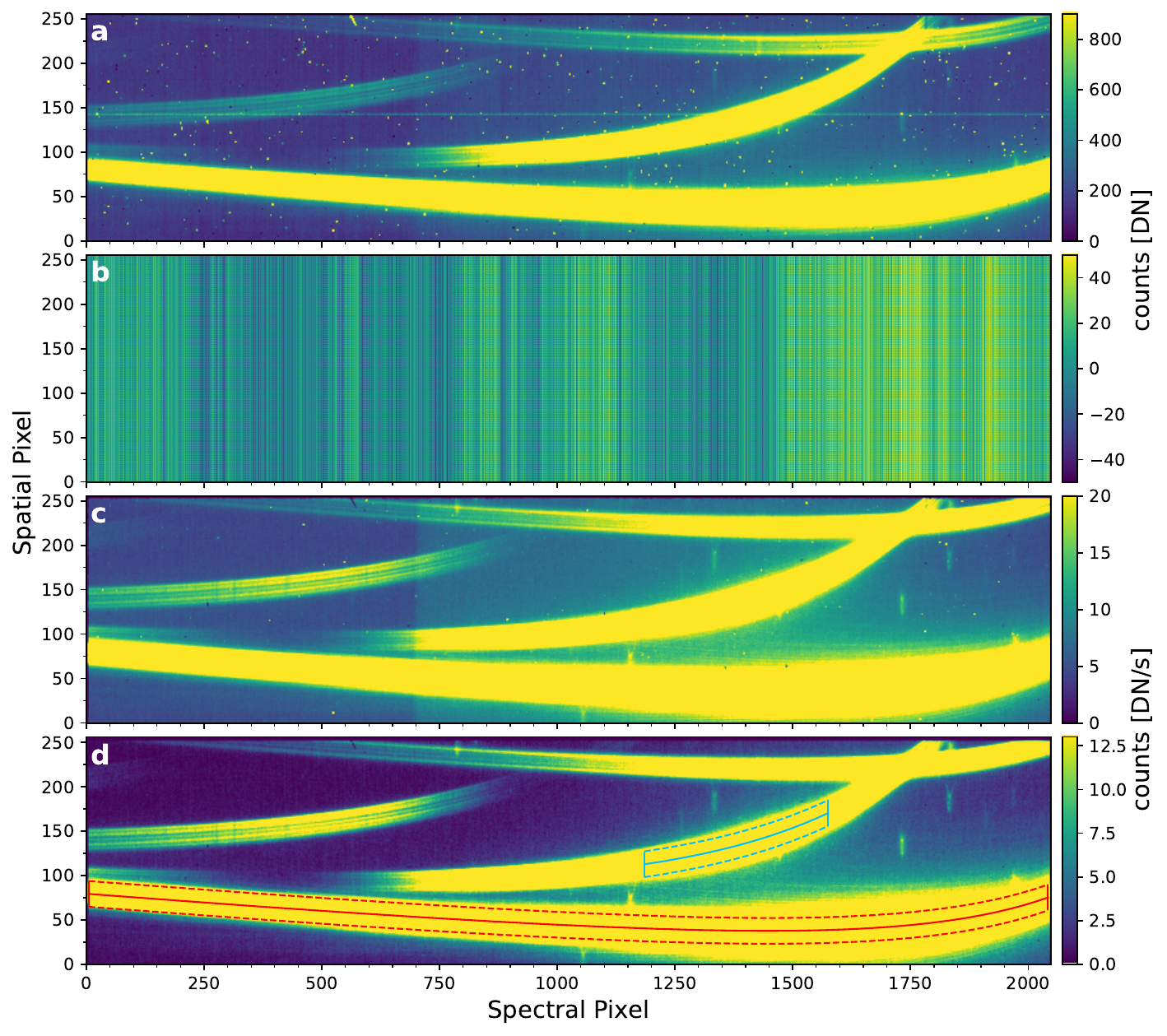}
    \caption{Data products at several stages of the reduction process for the NIRISS/SOSS eclipse observation of WASP-80\,b.
    \textbf{(a)} Last group of the 113$^{\rm th}$ integration, after superbias subtraction.
    \textbf{(b)} 1/$f$ noise in frame (a).
    \textbf{(c)} Integration 113 after ramp fitting and flat field correction.
    \textbf{(d)} Final calibrated frame after background subtraction and bad pixels interpolation. The full and dashed lines represent the center and edges, respectively, of the aperture boxes used to extract the first (red) and second (blue) order. The top trace is a contaminant from another star.
    \label{fig:Reduction_Steps}}
\end{figure*}

Most steps included in Stage 1 of \texttt{supreme-SPOON} come directly from the official \texttt{jwst} Data Management System (DMS) pipeline. However, we opted to skip the \texttt{RefPixStep} and perform a slightly modified version of the \texttt{OneOverFStep} from \texttt{supreme-SPOON}, as both steps serve the same purpose. The nominal \texttt{OneOverFStep} applied at the group level requires the subtraction of the zodiacal background beforehand to avoid rescaling it during the scaling of the group-wise median frame that will be subtracted from each image \citep{radica_awesome_2023}. However, in our case, where the variation in flux during the secondary eclipse is negligible compared to the background level in the wings of the SOSS point spread function (PSF), we simply subtracted a complete median frame for each group to reveal the 1/$f$ noise. By doing so, we avoid incorrect background subtraction, which could lead to inaccurate 1/$f$ noise removal. We then masked the traces, bad pixels and order 0 contaminants. To create the trace mask, we exploited the fact that photon noise increases in highly illuminated pixels. That is, we created a 2D detector map of the temporal standard deviations in flux using all median-subtracted images from the last group in an outlier-resistant manner, then created a mask with all pixels having a standard deviation higher than 5\% of the median value from the 2D map. Once the masks have been applied, we computed the value of the 1/$f$ noise by taking the median of each column, separately for even and odd rows \citep{feinstein_early_2023, radica_awesome_2023}. The resulting 1/$f$ noise, for which an example is shown in Figure \ref{fig:Reduction_Steps}(b), was then subtracted from each frame.

For the \texttt{JumpStep}, we applied the default algorithm from \texttt{jwst} with a rejection threshold of 5\,$\sigma$, as well as the jump step algorithm from \texttt{supreme-SPOON} with a rejection threshold of 10\,$\sigma$ as we find that some pixels are flagged in one algorithm but not in the other.

In Stage 1, we added the \texttt{ChargeMigrationStep} \citep{goudfrooij_algorithm_2024} from \texttt{jwst} because the signal reaches the flux threshold of $\sim$25,000\,ADU near the blaze peak in the order 1 trace \citep{albert_near_2023} during the last read for many pixels in the central region of the PSF, and even from group 3 for the brightest ones. Although the \texttt{ChargeMigrationStep} is primarily intended for imaging, it is also relevant for SOSS, as the effects of non-linearity are the same in both observing modes for bright objects. Groups flagged by this step were subsequently excluded from the slope calculations in the \texttt{RampFitStep} \citep{goudfrooij_algorithm_2024}.

We then proceeded with Stage 2 to perform corrections at the slope level. The \texttt{BackgroundStep} requires careful treatment because the optimal scaling of the Space Telescope Science Institute (STScI)\footnote{\url{https://jwst-docs.stsci.edu/jwst-calibration-pipeline-caveats/jwst-time-series-observations-pipeline-caveats/niriss-time-series-observations-pipeline-caveats\#NIRISSTimeSeriesObservationsPipelineCaveats-SOSSskybackground}} SOSS SUBSTRIP256 background model differs for each dataset (\citealp{radica_awesome_2023, fournier-tondreau_near-infrared_2024}). For our dataset, we also find that applying a small shift to the background model prior to its subtraction resulted in cleaner residuals around the pick-off mirror jump at column $\sim$700 (see Figure \ref{fig:Reduction_Steps}(c)). To compute the best shift value, we used the median row between the order 2 and order 3 traces by considering only the 75 columns on each side of the pick-off mirror jump for both the median data frame and the background model, and we scaled the data to match the background model level. We then minimized the residuals between the data and the background model shifted at different values, and found that the background model must be shifted left by 1.75 pixels.

The scaling of the background model also needs to be performed separately for each side of the background jump (\citealp{lim_atmospheric_2023, fournier-tondreau_near-infrared_2024}). To determine the factors for scaling the background model, we first had to consider that no region on the detector was completely non-illuminated due to the extended PSF wings from all traces. After masking some outliers, we first divided the median data frame by the shifted background model, and took the 10$^{\rm th}$ percentile of the ratios in a small region on the left side of the detector ($x \in [233,244],~ y \in [5,105]$) as the left scale factor. Since the right side was too illuminated to use the same method, we determined its scale factor by testing different values and selecting the one that best eliminated the background step. To verify this, we ensured that the slope in the central rows around the step, where the flux is minimal, was nearly zero after subtracting the scaled background, but slightly positive to account for the small increase in flux from the left to the right side of the step. Therefore, we determined that the best scaling factors redward ($x<\sim$700) and blueward ($x>\sim$700) of the background jump were 0.86 and 0.85, respectively. We ended Stage 2 by applying the \texttt{BadPixStep}, which uses the median of a surrounding box to interpolate hot, warm, as well as all \texttt{DO\_NOT\_USE} pixels. An example of the final product for one integration is shown in Figure \ref{fig:Reduction_Steps}(d).

We extracted the spectra with the \texttt{Extract1dStep} module of the \texttt{jwst} pipeline using a box width of 30 pixels. We did not decontaminate the order 1 and 2 traces due to the bright order 1 trace contaminant at the top of the detector, which prevented effective modeling of the traces. Additionally, the dilution caused by the overlap of the order 1 and order 2 traces at the longest wavelengths is negligible \citep{darveau-bernier_atoca_2022, radica_applesoss_2022}. We found a small deviation between the spatial positions of the target traces and those in the \texttt{spectrace} reference file\footnote{file \texttt{jwst\_niriss\_spectrace\_0023.fits} from \url{https://jwst-crds.stsci.edu/}}, which could reach 4 pixels at the red end of both traces. Therefore, we adjusted the $y$ positions of the traces for the extraction. Finally, we clipped all data points that deviated by more than 5\,$\sigma$ in time and used the wavelength solution computed by the \texttt{PASTASOSS} package \citep{baines_characterization_2023} using the position of the pupil wheel during the observation of WASP-80\,b. Figure \ref{fig:Reduction_Steps} shows a summary of the major reduction steps, including the box aperture used for extraction in both traces.

We also reduced the data using another independent pipeline to assess the self-consistency of our results, which is described in the Appendix (\S\ref{sec:add_reduc}).\\

\section{Light Curve Fitting}
\label{sec:Fitting}

\subsection{White-light curve fitting}
\label{sec:wlc_fit}

Before performing the light curve analysis, we masked all wavelengths where contaminants overlap with the traces on the detector, as they could dilute the eclipse signal. We masked two 0$^{{\rm th}}$ order contaminants of field stars: one in the order 1 trace affecting wavelengths between 0.905 and 0.919\,$\mu$m ($x\in [1964,1980]$), and one in the order 2 trace between 0.716 and 0.722\,$\mu$m ($x\in [1465,1479]$). There is also an order 1 contaminant from another source at the top of the detector. To avoid it, we cut the order 2 trace at column 1575, where the lower wing of the contaminant trace has a negligible contribution to the order 2 flux. Cutting this portion of the order 2 trace prevented us from using the light curves between 0.600 and 0.674\,$\mu$m. However, the impact on the results is minor considering the large bins of the spectrum, as well as the decreasing throughput and signal-to-noise ratio at those wavelengths. The blue lines in Figure \ref{fig:Reduction_Steps}d show the portion of order 2 that was used in the final spectrum.

We constructed a white-light curve for order 1 (0.850--2.827\,$\mu$m) and order 2 (0.674--0.850\,$\mu$m) by summing the normalized light curves (which allows for a comparison between different instruments) from all wavelengths for each order. Although the fitting method is the same for both orders, we mainly focus on the order 1 white-light curve in what follows, as it has higher signal-to-noise ratio than order 2 and order 1 represents the majority of the final spectrum.

\begin{figure}
	\centering
	\includegraphics[width=\columnwidth]{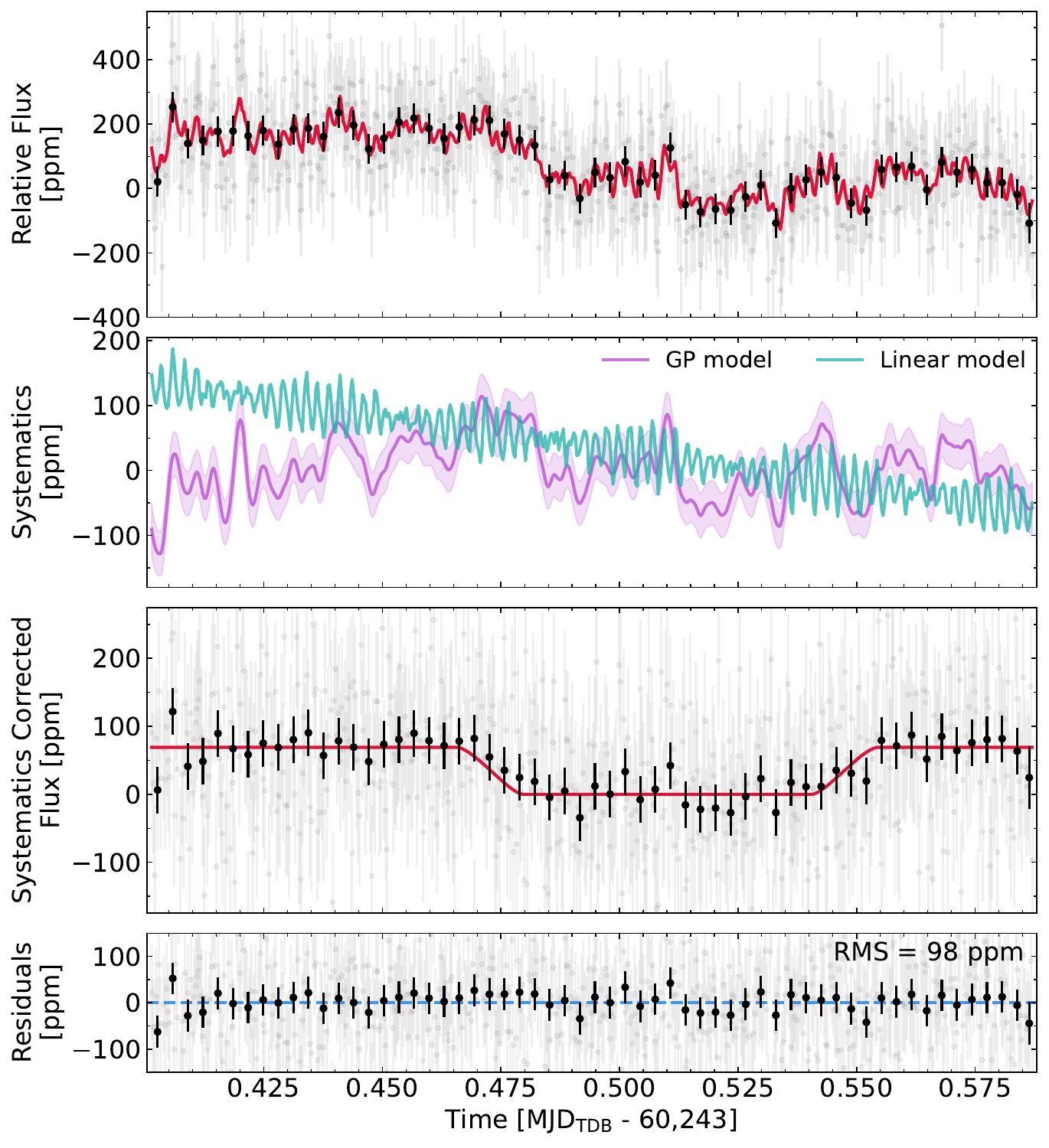}
    \caption{\textbf{Top:} Raw NIRISS/SOSS order 1 (0.85--2.83\,$\mu$m) white-light curve with 1-$\sigma$ error bars (gray), representing the standard deviation on the out-of-eclipse data points, and its temporally binned points (black). The best-fit model, including the eclipse and the systematics, is overlapped in red. \textbf{Second:} Systematics removed from the white-light curve. The linear systematics model from equation \ref{eq:lin_sys} is shown in turquoise and the remaining correlated noise modeled with the GP along with its 1-$\sigma$ confidence interval is shown in purple. \textbf{Third:} Detrended order 1 white-light curve (gray, binned points in black) once the systematics of the second panel are removed from the data of the top panel along with the eclipse model (red). The error bars represent the best-fit scatter parameter ($\sigma=102.9$\,ppm). As more easily seen on the binned points, the error bars seem overestimated as expected from our Allan variance plot from Figure \ref{fig:allan}, which is explained by potential remaining correlated noise at the integration level. \textbf{Bottom:} Residuals to the best-fit model and their RMS value (based on the gray unbinned points).
    \label{fig:wlc_fit}}
\end{figure}

A tilt event occurred around integration 84, approximately 0.63\,hour after the beginning of the observation. Tilt events are caused by small, abrupt changes in the position of one of the primary mirror segments of the telescope, resulting in a change in the SOSS PSF that creates a jump in the light curves \citep{alderson_early_2023, rigby_science_2023,coulombe_broadband_2023}. We decided to simply discard the first 83 integrations, as they represent a small fraction of the baseline. The resulting order 1 white-light curve is shown in the top panel of Figure \ref{fig:wlc_fit} and that of order 2 is shown in the top left panel of Figure \ref{fig:o1below15}.

A comparison between our order 1 white-light curve from the \texttt{supreme-SPOON} reduction and the one from the additional reduction is shown in Figure \ref{fig:2wlcs} of the Appendix. The light curves reveal the presence of significant systematics that are independent of the reduction since they follow similar trends. These systematics therefore come from the instrument and the host star, and we correct them during the eclipse model fitting. Our complete model is described as follows:

\begin{equation}
\label{eq:full_model}
m(t) = F(t) \cdot L(t) + C(t) \,,
\end{equation}
where $F(t)$ is the eclipse model, $L(t)$ is the linear systematics model, and $C(t)$ is the model of the remaining correlated noise.

Our linear systematics model is defined by three components:

\begin{equation}
\label{eq:lin_sys}
L(t) = b(t-t_{\rm med}) + c\lambda_\mathrm{PC_3} + d + 1 \,.
\end{equation}
The first term is a linear trend with a slope $b$, centered at the median time $t_{\rm med}$ of the TSO. It considers any systematic varying on a long-term basis. The second term of $L(t)$ is a common trend found in every NIRISS/SOSS TSO \citep{albert_near_2023,coulombe_broadband_2023,radica_muted_2024,Coulombe2025highlyreflectivewhiteclouds}, caused by temperature variations in the instrument electronics compartment that induce small oscillations in the data \citep{mcelwain_james_2023, albert_near_2023}. To retrieve this trend, we performed a temporal principal component analysis (PCA), which can track morphological changes in the traces, following a similar procedure as \cite{coulombe_broadband_2023} and  \citet{Coulombe2025highlyreflectivewhiteclouds}. As shown in Figure \ref{fig:PCA} of the Appendix, these oscillations appear in the third principal component and translate into changes in the full width at half maximum (FWHM) of the traces. We used this third principal component ($\lambda_{\rm PC_3}$), standardized and scaled by the coefficient $c$, to detrend our white-light curve. Beforehand, we subtracted a uniform filter of size 151 pixels to the eigenvalue time series to keep only the high-frequency signal of this oscillation. We did not use the first principal component of the PCA to detrend our white-light curve because it predominantly contains the eclipse signal. We also excluded the second eigenvalue as we find that it does not improve the fit ($\Delta{\rm log}\mathcal{Z}=0.14\pm0.27$). The third term in our linear systematics model, $d$, is a normalization parameter.

After detrending against these systematics, we observed remaining time-correlated signals in the residuals which cannot be attributed to the planet. These residuals do not exhibit the same shape at each wavelength and may originate from a combination of both stellar and instrumental effects. To account for these systematic residuals, we employed a Gaussian Process (GP) with a Matérn 3/2 kernel, which we computed using the \texttt{celerite} open-source python package \citep{foreman-mackey_fast_2017}.

We simultaneously fitted all components from Equation \ref{eq:full_model} to our order 1 white-light curve. For the eclipse model $F(t)$, we used the \texttt{batman} package \citep{kreidberg_batman_2015} and considered a circular orbit ($e=0$), consistent with the results from \cite{bell_methane_2023}. Using the value from \cite{triaud_wasp-80b_2015}, we fixed the planet-to-star radius ratio at $R_{\rm p}/R_*=0.17137$. Based on the results from \cite{bell_methane_2023}, we fixed the orbital period to $P=3.067851945$\,days, and put Gaussian priors on the scaled semi-major axis $a/R_*$ ($\mathcal{N}[12.643, 0.032]$) and the inclination $i$ ($\mathcal{N}[88.938, 0.059]^\circ$) (see Table \ref{tab:1} for the list of system parameters used in this analysis).

When allowing for the time of mid-eclipse to vary, we obtained a stronger eclipse detection ($\Delta{\rm log}\mathcal{Z}=2.33\pm0.27$), but measured a mid-eclipse time delayed by $12.8\pm4.6$ minutes from the expected timing. This discrepancy could suggest an eccentricity of $0.065\pm0.023$, but is inconsistent with the results of zero eccentricity from \cite{triaud_wasp-80b_2015} and \cite{bell_methane_2023}. In our case, this delay could result from the significant systematics present in the light curve. We therefore opted to fix the time of mid-eclipse to the expected value from a circular orbit with a mid-transit time of $T_0=56487.925006$\,MJD \citep{bell_methane_2023}, corrected for the light-travel time between superior and inferior conjunctions.

To avoid bias, we allowed for negative values in the eclipse depth $F_{\rm p}/F_*$ by using a wide uninformative prior ($\mathcal{U}[-500, 500]$\,ppm). We fitted for a scatter parameter $\sigma$ to represent the uncertainties and applied a wide uniform prior to it ($\mathcal{U}[0, 500]$\,ppm). The fit also included three systematics parameters: the linear slope $b$ ($\mathcal{U}[-1,1]$), the coefficient of the third principal component $c$ ($\mathcal{U}[-1, 1]$), and the normalization parameter $d$ ($\mathcal{U}[-250, 250]$\,ppm). We finally fitted for the GP parameters, which are the amplitude $a_{\rm GP}$ ($\mathcal{LU}[1, 200]$\,ppm) and the length-scale $\rho_{\rm GP}$ ($\mathcal{LU}[1, 60]$\,min). The prior on the length scale was chosen such that the GP did not model the scatter between individual points, but modeled the visible correlated trends that vary on length scales no longer than one hour. Our complete model therefore includes 9 free parameters that we fitted using dynamic nested sampling \citep{higson_dynamic_2019} with the \texttt{dynesty} package \citep{speagle_dynesty_2020, koposov_joshspeagledynesty_2023}. We initialized the baseline run with 450 live points and a $\Delta\mathrm{ln}\mathcal{Z}$ tolerance of 0.05, and we allowed for an infinite number of iterations.

The priors and results of the order 1 white-light curve fits for both reductions are detailed in Table \ref{tab:4}. Figure \ref{fig:wlc_fit} shows the order 1 white-light curve from the \texttt{supreme-SPOON} reduction along with its best-fit model and the systematics-corrected white-light curve. The residuals follow the Poisson noise trend (1/$\sqrt{N}$, where $N$ is the number of integrations) at bin sizes larger than $\sim300$ s, as shown in Figure \ref{fig:allan}. We explain the steeper slope at bin sizes $\lesssim300$ s by the presence of remaining correlated noise at frequencies that are not targeted by the GP.

\subsection{Spectrophotometric light curve fitting}
\label{sec:spectro_fit}

\begin{figure*}
	\centering
	\includegraphics[width=0.9\textwidth]{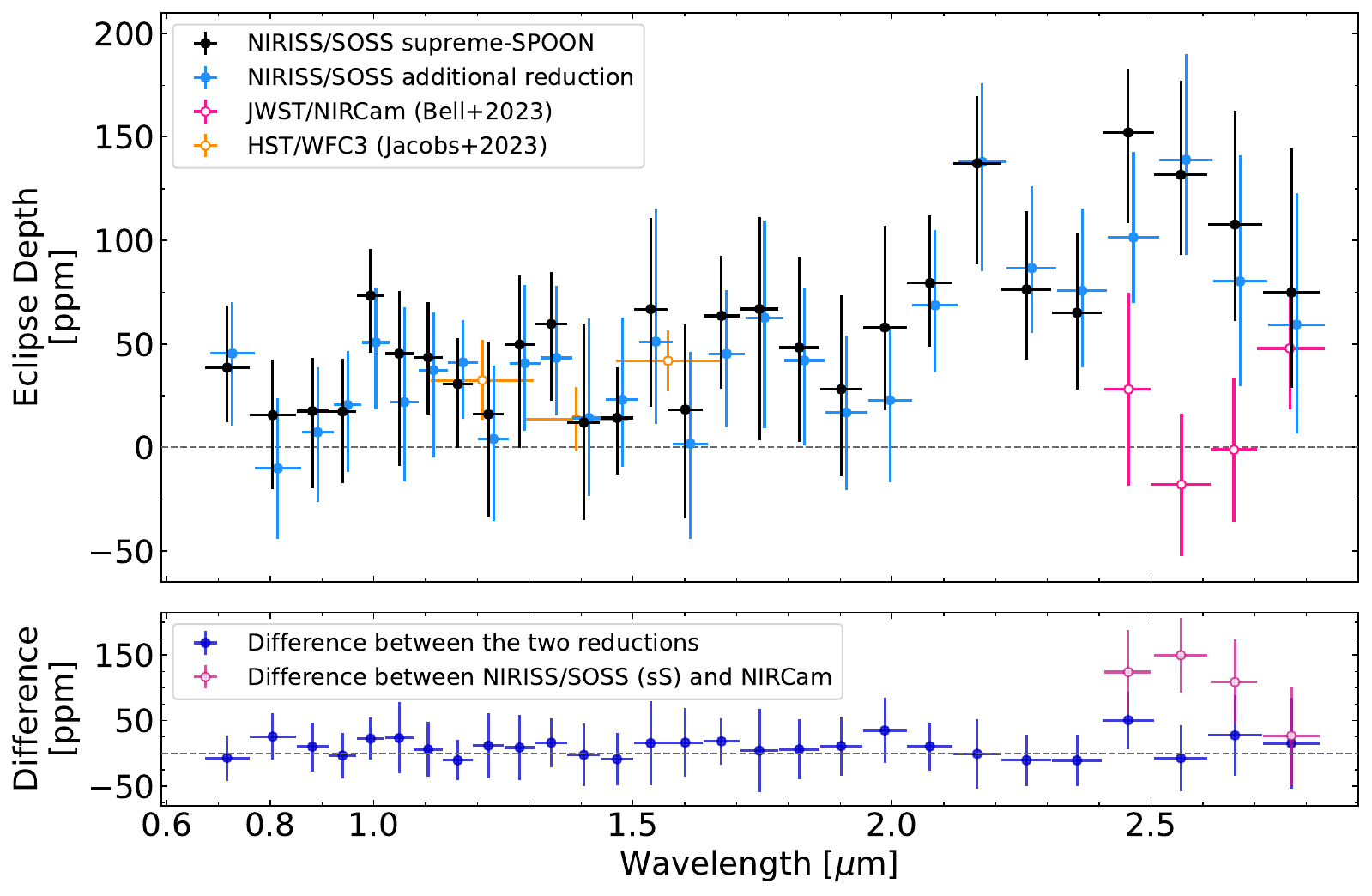}
    \caption{\textbf{Top:} Eclipse spectrum of WASP-80\,b. The black points with the 1-$\sigma$ vertical error bars show the NIRISS/SOSS spectrum obtained with the \texttt{supreme-SPOON} reduction pipeline and the blue points show the spectrum obtained with the additional reduction, shifted by 0.01\,$\mu$m for clarity. The bins from the order 2 trace are the first 2 points of this spectrum and the other bins come from the order 1 trace. The HST/WFC3 data points from \cite{jacobs_probing_2023} are shown in orange as well as the first part ($2.42-2.81$ $\mu$m) of the JWST/NIRCam spectrum from \cite{bell_methane_2023} in pink, binned down to the NIRISS/SOSS resolution. \textbf{Bottom:} The difference in SOSS spectra between the \texttt{supreme-SPOON} and other pipeline is shown in dark blue, where the vertical error bars represent the maximum ones of the two reductions for each bin. The difference between the SOSS \texttt{supreme-SPOON} and NIRCam spectra is shown in dark pink, where the vertical error bars represent the square root of the sum of the squares of the individual error bars.
    \label{fig:spectra}}
\end{figure*}

We then proceeded with fitting the spectrophotometric light curves of both order 1 and order 2. We fixed the time of mid-eclipse to the expected value, and the scaled semi-major axis and the inclination to their values from the model of the order 1 white-light curve with the highest likelihood (see Table \ref{tab:4}). Since the order 2 light curves present similar systematics as that of order 1, the same methods described in the previous section were also applied to the order 2 light curves. Given this presence of significant systematics in the data, we binned the light curves to an average resolving power of $R\sim22$ for order 1 and $R\sim9$ for order 2 to enhance the signal-to-noise ratio of each bin. This results in a spectrum comprising 26 bins from order 1 and 2 bins from order 2.

For the spectroscopic fit, we narrowed some priors to facilitate convergence for each bin while ensuring the 99.7\% confidence interval of each parameter from the white-light curve fit remained within the prior ranges. For each bin, we fitted for the linear slope $b$ ($\mathcal{U}[-5\times10^{-3}, 10^{-4}]$), the coefficient $c$ of the third eigenvalue from the PCA ($\mathcal{U}[-10^{-4}, 10^{-3}]$), the normalization parameter $d$ ($\mathcal{U}[-50, 150]$ ppm), the scatter parameter $\sigma$ ($\mathcal{U}[0, 1000]$\,ppm) and the eclipse depth $F_\mathrm{p}/F_*$ ($\mathcal{U}[-100, 400]$\,ppm). As mentioned earlier, the correlated noise $C(t)$ is chromatic, which is why we allowed the GP parameters to vary for each bin. Thus, we also fitted for the GP amplitude $a_{\rm GP}$ ($\mathcal{LU}[1, 150]$\,ppm) and the length-scale $\rho_{\rm GP}$ ($\mathcal{LU}[1, 60]$\,min). The spectrophotometric fit therefore had 7 free parameters for each bin, which were fitted using the same nested sampling method as for the white-light curve.

The final eclipse depth of each bin was determined by taking the maximum of the posterior distribution and the associated 1-$\sigma$ confidence region of the highest posterior density interval. We proceeded in the same way for the spectrophotometric fit with the additional, independent data reduction, except using the highest likelihood values of the scaled semi-major axis and inclination from its own order 1 white-light curve fit (see Table \ref{tab:4}). The final eclipse spectra are presented in Figure \ref{fig:spectra} for both reductions, where they are also compared to the HST/WFC3 data from \cite{jacobs_probing_2023} and the first part ($2.42-2.81$ $\mu$m) of the JWST/NIRCam spectrum from \cite{bell_methane_2023}.\\

\section{Atmosphere modeling}
\label{sec:modeling}

We interpret our NIRISS/SOSS secondary eclipse spectrum of WASP-80\,b using a range of atmospheric forward models and retrievals to simulate the contributions of reflected light and thermal emission to the measured planetary flux, as presented in the subsections below.

\subsection{SCARLET atmospheric retrievals}
\label{sec:scarlet_ret}

We perform atmospheric retrievals on our WASP-80\,b secondary eclipse spectrum using the SCARLET framework \citep{benneke_atmospheric_2012,Benneke_2013,benneke2015strictupperlimitscarbontooxygen,Benneke_2019,Pelletier_2021,coulombe_broadband_2023,Roy_2023,benneke2024jwstrevealsch4co2,Piaulet_Ghorayeb_2024} and following a methodology similar to that presented in \citet{Coulombe2025highlyreflectivewhiteclouds}. SCARLET takes as input a set of altitude-dependent molecular abundances, a T-P profile, a cloud top deck pressure, and a reflected light geometric albedo to produce a forward model of the planetary flux. The algorithm then uses the affine invariant Markov chain Monte Carlo ensemble sampler \texttt{emcee} \citep{emcee2013} to explore the parameter space. The models are simulated at a fixed resolving power of $R =$\,15,625 and subsequently binned to the resolution of the data assuming a uniform throughput. The modeled planetary flux is converted to units of planet-to-star flux ratio considering a PHOENIX model \citep{husser_new_2013} interpolated to the stellar parameters of WASP-80 ($T_\mathrm{eff,*} = 4145$\,K, $\log~g$ = 4.6, [Fe/H] = -0.14, \citealt{triaud_wasp-80b_2013}) for the stellar spectrum. We perform one retrieval assuming equilibrium chemistry and one where the molecular abundances are free and considered constant with altitude (i.e., free chemistry).

For the chemical equilibrium retrieval, we consider the following species: H$_2$, H, H- \citep{Bell1987,John1988}, He, H$_2$O \citep{Polyansky_2018}, CH$_4$ \citep{hargreaves_accurate_2020}, CO$_2$ \citep{yurchenko_exomol_2020}, CO \citep{li_rovibrational_2015}, OH \citep{hitemp2010}, Na \citep{allard_new_2019}, K \citep{allard_kh2_2016}, C$_2$H$_2$ \citep{chubb_exomol_2020}, NH$_3$ \citep{Coles2019}, HCN \citep{Barber_2013}, H$_2$S \citep{chubb_marvel_2018}, and PH$_3$ \citep{sousa-silva_exomol_2015}. The abundances of these species are interpolated in temperature, pressure, metallicity ([M/H]), and carbon-to-oxygen ratio (C/O) from a grid of chemical equilibrium abundances produced using FastChem2 \citep{Stock_2022}.

We model the T-P profile using a free parametrization \citep{Pelletier_2021,Pelletier_2023,coulombe_broadband_2023,Bazinet_2024} and considering 10 temperature points ($\mathcal{U}[100,4400]$\,K) that are equally spaced in log-pressure ($P= 10^{2}$--$10^{-7}$\,bar). The 10 T-P profile points are interpolated to the 40 layers of the atmosphere considered for the forward models using a spline interpolation. Following the method described in \citet{Pelletier_2021}, we penalize against the second derivative of the T-P profile using a smoothing hyperparameter value of $\sigma_s=100$\,K/dex$^2$ to avoid unphysical temperature variations over short pressure length scales.

We also let free the atmospheric metallicity ([M/H], $\mathcal{U}[-1,3]$), carbon-to-oxygen ratio (C/O, $\mathcal{U}[0,1]$), cloud top pressure ($\log P_\mathrm{Cloud}$, $\mathcal{U}[-8,2]$\,$\log$[bar]), and reflected light geometric albedo ($A_\mathrm{g}$, $\mathcal{U}[-1,1]$). The geometric albedo is constant with wavelength and is multiplied by the squared ratio of the wavelength-dependent planetary radius $R_\mathrm{p}(\lambda)$ to the semi-major axis $a$ to produce the contribution of reflected light to the planetary flux ($F_\mathrm{p}/F_\mathrm{s}= (R_\mathrm{p}(\lambda)/a)^2)A_\mathrm{g}$). We explore the parameter space for 30,000 steps and discard the first 60\% as burn-in to produce the posterior distributions of the parameters.

We also perform a free chemistry retrieval where, instead of the molecular abundances being dictated by the atmospheric metallicity and carbon-to-oxygen ratio, we assume the abundances of species to be constant with altitude and directly fit for their values. As in \citet{bell_methane_2023}, we fit for the volume mixing ratio of water, methane, carbon monoxide, carbon dioxide, and ammonia. We do not fit for sulfur dioxide as its main opacity bands are found at 4.05, 7.7, and 8.5\,$\mu$m \citep{Tsai_2023,Powell_2024}, outside the domain covered by our observations. We assume wide uniform priors in log-space for the abundances ($\mathcal{U}$[-15,0]).
We repeat the same procedure as for the chemical equilibrium retrieval and explore the parameter space for 30,000 steps, again discarding the first 18,000 steps as burn-in.

\subsection{\texttt{PICASO}/\texttt{VIRGA} cloud modeling}
\label{sec:cloud_modeling}

We use the \texttt{PICASO} \citep{batalha_exoplanet_2019} and \texttt{VIRGA} \citep{Rooney2022} python packages to produce reflected light models across the NIRISS/SOSS wavelength range for a variety of cloud species expected to condense at the temperatures of WASP-80\,b. \texttt{VIRGA} takes as input a T-P profile, an atmospheric metallicity ([M/H] = 0, \texttt{VIRGA} cannot handle non-solar metallicities due to chemical limitations), and a mean-molecular weight ($\mu_\mathrm{m}$ = 2.3). It then returns a list of cloud species whose condensation curves cross the given T-P profile at some altitude in the atmosphere. We provide to \texttt{VIRGA} 1D self-consistent temperature profiles produced with SCARLET assuming an atmosphere metallicity of 6 times solar, a carbon-to-oxygen ratio of 0.35 \citep{bell_methane_2023}, Bond albedo values of $A_\mathrm{B} = 0$ and 0.25, and a heat redistribution factor of $f$ = 1 ($f$ = 1 corresponds to full redistribution and $f$ = 2 to no redistribution with a uniform dayside). We note that, despite the fact that \texttt{VIRGA} can only produce cloud models for solar metallicity atmospheres, we still supply a T-P profile from a 6 times solar metallicity atmosphere as it is likely a more accurate representation of the planet's thermal structure.

The condensates recommended by \texttt{VIRGA} are Cr[s], KCl, Mg$_2$SiO$_4$, MgSiO$_3$, MnS, Na$_2$S, and ZnS. We use \texttt{PICASO} to produce the reflected light model for a specific condensate species and assume a given vertical mixing coefficient $K_{zz}$ and sedimentation efficiency $f_\mathrm{sed}$. For the mixing coefficient, we follow the relation presented in \citet{Moses_2021} and consider a pressure-dependent mixing coefficient ($K_{zz} = 5\times10^{8} P^{-0.5}[H/620\,\mathrm{km}][T_\mathrm{eq}/1450\,\mathrm{K}]^4$), where $H$ = 232\,km is the atmospheric scale height of WASP-80\,b. As for the sedimentation efficiency, we run the models considering values of 0.1 and 3 \citep{webber_effect_2015}, as these are generally the values expected in hot Jupiters and our own Jupiter, respectively. This range of values thus most likely covers the true sedimentation efficiency of the clouds on WASP-80\,b. \\

\section{Results and Discussion}
\label{sec:results_discussion}

\subsection{Eclipse depth measurements}
\label{sec:eclipse_depths}

For the broadband order 1 light curve (0.85 to 2.83\,$\mu$m) presented in Figure \ref{fig:wlc_fit}, we retrieve an eclipse depth of $65_{-35}^{+28}$\,ppm using the fit method from Equation \ref{eq:full_model}, corresponding to a 1.9-$\sigma$ measurement. By comparing the evidence of our best-fit model with the one from a fit using a noise-only model, we also obtain a preference of $\sim$2\,$\sigma$ for our model including the eclipse ($\Delta\mathrm{log}\mathcal{Z}=0.98$, \citet{Benneke_2013}).
When we do not use a GP model in the fit, keeping everything else the same, we obtain a consistent eclipse depth of $70_{-11}^{+10}$\,ppm, yielding a higher detection significance of $6.4\,\sigma$. As shown by the increased error bars, the inclusion of the GP in the light curve modeling seems to adequately capture the additional uncertainties associated with the systematic noise in the data and thus provides a more realistic uncertainty on the eclipse depth.
For the broadband order 2 light curve (0.674 to 0.850\,$\mu$m) fit, presented in the left panels of Figure \ref{fig:o1below15} of the Appendix and performed using the same method as described in \S\ref{sec:wlc_fit}, the retrieved eclipse depth is $26_{-37}^{+27}$\,ppm. This inconclusive result was expected given the important systematics and low signal-to-noise ratio in order 2.

In the eclipse spectrum shown in Figure \ref{fig:spectra}, some error bars are larger than others and occasionally cross the zero eclipse depth line, particularly in the bluer part of the spectrum. We notice that when the GP parameters are more tightly constrained in a certain bin, the error bars on the eclipse depth become larger. Again, this is because we can better account for the uncertainties on the correlated noise when we can model it better. Moreover, between 1.1 and 1.7\,$\mu$m, our results agree well with those from \cite{jacobs_probing_2023} obtained with HST/WFC3.

\begin{figure*}
	\centering
	\includegraphics[width=\textwidth]{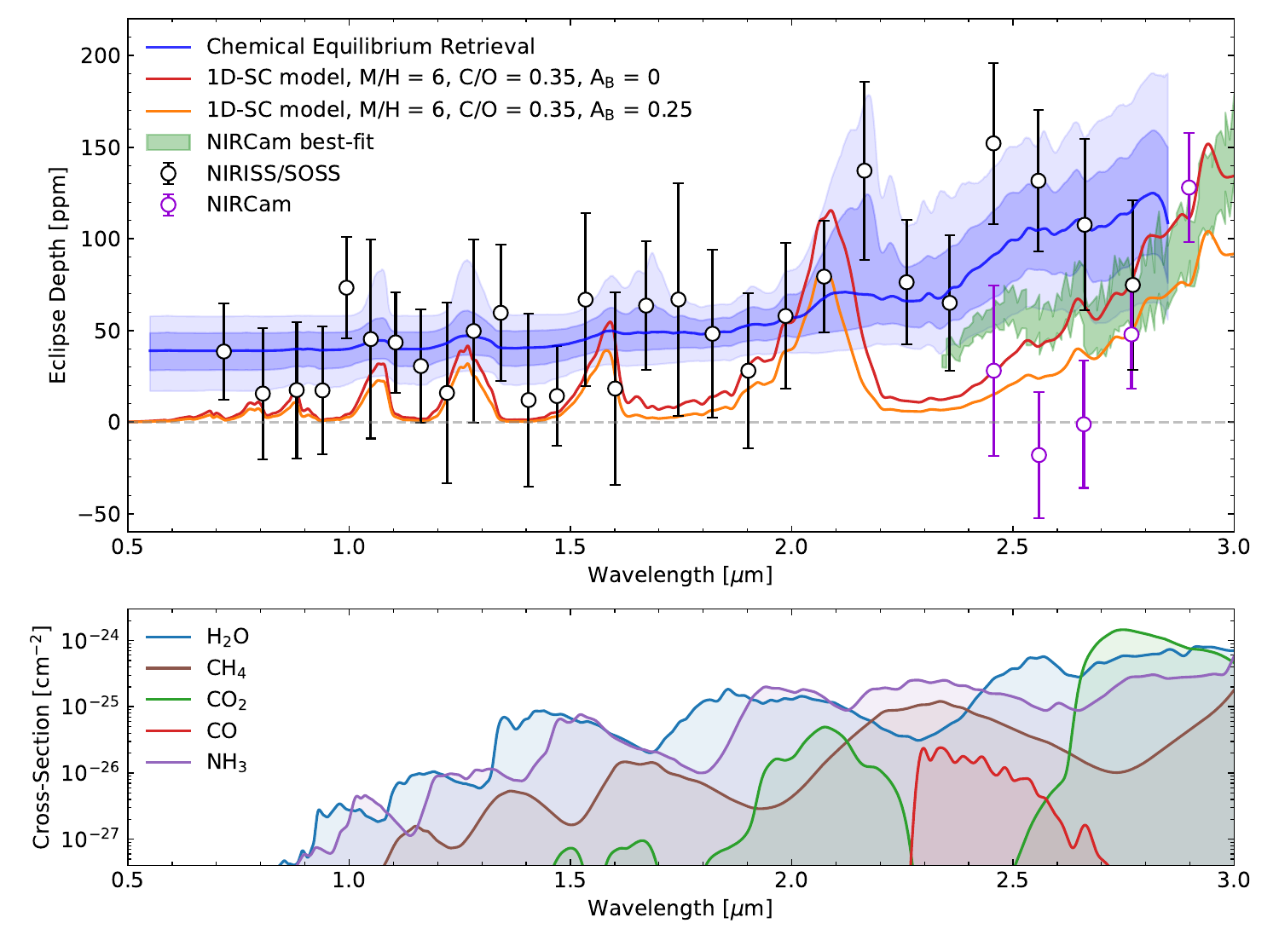}
    \caption{\textbf{Top:} Median model (blue) along with the 1- and 2-$\sigma$ confidence intervals from the SCARLET chemical equilibrium retrieval to the secondary eclipse spectrum of WASP-80\,b (black). The inclusion of a reflected light geometric albedo in this retrieval leads to a contribution from reflected light of $\sim$30\,ppm to explain the data at short wavelengths. For comparison, we show  1D self-consistent models, assuming Bond albedo values of 0 (red) and 0.25 (orange), produced with SCARLET for an atmospheric metallicity of 6 times solar and carbon-to-oxygen ratio of 0.35. We show in green the 1-$\sigma$ confidence region of the best-fit model to the NIRCam secondary eclipse spectrum (purple, binned down to the NIRISS/SOSS resolution) as presented in \citet{bell_methane_2023}. \textbf{Bottom:} Cross-sections of water (blue), methane (brown), carbon dioxide (green), carbon monoxide (red), and ammonia (purple); the species considered in the free chemistry retrieval, over the NIRISS/SOSS wavelength range. The cross-sections, shown at a fixed resolving power of $R=100$, are computed at a temperature of 800\,K and pressure of 10\,mbar, roughly corresponding to the photospheric conditions of WASP-80\,b. Increasing cross-sections correspond to stronger absorption, and thus, lower thermal emission.
    \label{fig:spec_retrieval}}
\end{figure*}

Below 2\,$\mu$m, the spectrum is mostly flat and shows hints of excess flux above what is expected from thermal emission alone given the temperature of WASP-80\,b. Although the NIRISS/SOSS eclipse depths are mostly consistent with the 1D self-consistent models in Figure \ref{fig:spec_retrieval} within their 1-$\sigma$ uncertainties, they are almost always systematically higher. To better visualize this result, we performed a light curve fit on a large bin by summing all order 1 light curves below 1.5\,$\mu$m (0.85--1.50\,$\mu$m), where the thermal emission is negligible ($\sim$2\,ppm on average). The result is shown in the right column of Figure \ref{fig:o1below15}. The measured eclipse depth for this bin is $34_{-28}^{+25}$\,ppm, corresponding to a measurement significance of 1.2\,$\sigma$.  

We also observe a rise in the redder part of the spectrum ($\lambda\gtrsim2$\,$\mu$m), where the eclipse depth detections are more robust. Our depths in this spectral range are larger than those from the previous observations of WASP-80\,b with JWST/NIRCam analyzed by \cite{bell_methane_2023}, as shown in Figure \ref{fig:spectra}. Between 2.4 and 2.7\,$\mu$m, we detect significant eclipse depths, whereas the binned values of \cite{bell_methane_2023} are consistent with 0 within 1\,$\sigma$.
To quantify our detection of the planetary flux at these wavelengths, we performed the eclipse fit on a larger bin with a higher signal-to-noise ratio than each bin of our spectrum, by summing all light curves between 2.40 and 2.70\,$\mu$m and using the same detrending method as detailed in \S\ref{sec:wlc_fit}. As expected, the eclipse is deeper than that of the order 1 white-light curve, with an eclipse depth of $128_{-34}^{+31}$\,ppm, corresponding to a detection significance of 3.8\,$\sigma$ in that wavelength range. The discrepancy between the NIRISS and NIRCam data is discussed further in \S\ref{sec:NIRISS_vs_NIRCam}.

\subsection{Effects related to light curve modeling}
\label{sec:effects_lc_modeling}

To assess the robustness of our light curve modeling approach, we computed a spectrum without using a GP model. We observe a similar effect as with the white-light curve: the eclipse depths are consistent with the spectrum that includes a GP model in the light curve fit, but the error bars on each bin are smaller. More precisely, the 1-$\sigma$ upper limit of each bin is almost equal in both spectra, whereas the lower limits are higher in the spectrum that does not include a GP model.

We also obtained a spectrum by fixing the time of mid-eclipse to the value retrieved from the white-light curve fit in which it was free to vary. The resulting spectrum has mostly the same features as our reference one but is $\sim$30\,ppm higher on average, which was expected as the fit favors a different baseline. However, it would lead to unexpectedly high eclipse depths compared to previous studies and theoretical models, as well as an unexpected eccentricity as discussed in \S\ref{sec:wlc_fit}.
This spectrum still remains consistent with the reference spectrum within their 1-$\sigma$ confidence intervals.

The comparison between the spectra obtained from these different modeling approaches is presented in Figure \ref{fig:effects_model_spectra}.

\subsection{Results from the atmospheric retrieval}
\label{sec:results_retrievals}

Our fit from the chemical equilibrium retrieval on the NIRISS/SOSS secondary eclipse spectrum of WASP-80\,b is shown in Figure \ref{fig:spec_retrieval}. We find that, when a reflected light geometric albedo is included in the retrieval, the short-wavelength planetary flux is relatively well explained by a constant contribution from reflected light of approximately 30\,ppm, which could be indicative of the presence of clouds. Towards the longer wavelengths, past 2\,$\mu$m, the gradual increase in the flux is due to the onset of thermal emission.

\begin{figure*}
	\centering
	\includegraphics[width=\textwidth]{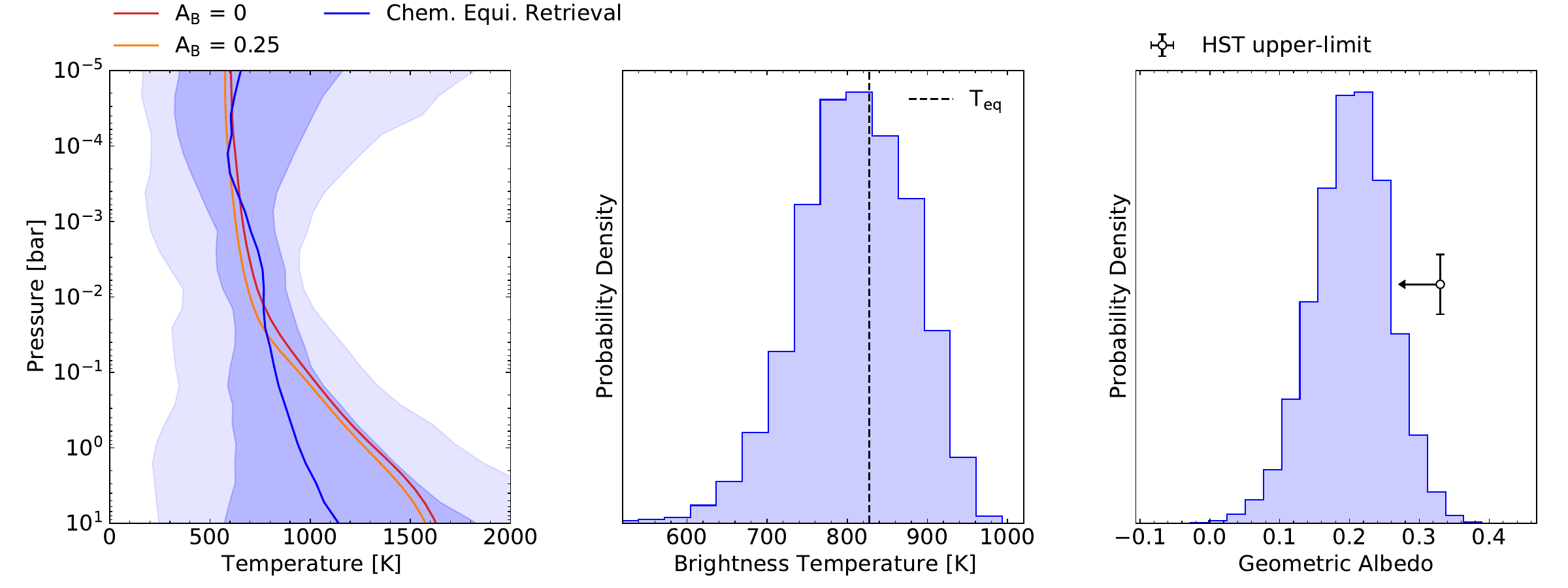}
    \caption{\textbf{Left:} Median temperature profile from the chemical equilibrium retrieval, along with its 1- and 2-$\sigma$ confidence regions. Our constraints show a slight preference for a non-inverted profile over the photosphere ($P$ = 10$^{-1}$--10$^{-2}$\,bar). The temperature profiles from the 1D self-consistent models, shown in red ($A_\mathrm{B}$ = 0) and orange ($A_\mathrm{B}$ = 0.25), are consistent within 1\,$\sigma$ of our measured profile. \textbf{Center:} Constraint on the brightness temperature ($T_\mathrm{B}$ = $811_{-70}^{+69}$\,K) as measured from the samples of the thermal emission spectra obtained from the retrieval. The equilibrium temperature of WASP-80\,b ($T_\mathrm{eq}$ = $824\pm19$\,K) assuming a Bond albedo of 0 and full heat redistribution ($f$ = 1) is shown for comparison. \textbf{Right:} Constraint on the reflected light geometric albedo ($A_\mathrm{g}=0.204^{+0.051}_{-0.056}$), which is constrained to 3.6\,$\sigma$ above 0. Our measured reflected light geometric albedo is consistent with the 3-$\sigma$ upper limit from HST ($A_\mathrm{g}<0.33$, \citealt{jacobs_probing_2023}).
    \label{fig:retrieval_constraints}}
\end{figure*}

Our retrieved T-P profile from the chemical equilibrium retrieval, as well as our constraints on the brightness temperature and reflected light geometric albedo of the planet are shown in Figure \ref{fig:retrieval_constraints}. We find that the T-P profile is consistent with those from our self-consistent models. There is a preference for non-inverted scenarios over the photosphere ($P=10^{-1}$--$10^{-2}$\,bar), which results in tentative molecular absorption features in the thermal portion of the planetary spectrum (Figure \ref{fig:spec_retrieval}), consistent with expectations for planets in this equilibrium temperature regime \citep{Baxter_2020}. We also constrain the brightness temperature of the planet over the NIRISS/SOSS bandpass by first integrating the samples of the thermal emission spectra from 1.5 to 2.83\,$\mu$m, weighted by the stellar spectrum and the throughput of the instrument. We then compute the corresponding blackbody temperature that would produce the same value of the bandpass-integrated flux. Our measured value of dayside brightness temperature is $T_\mathrm{B}=811_{-70}^{+69}$\,K.

We obtain a constraint on the reflected light geometric albedo of $A_\mathrm{g}=0.204_{-0.056}^{+0.051}$, in agreement with the 3-$\sigma$ upper limit of $A_\mathrm{g}<0.33$ derived from HST/WFC3 by \cite{jacobs_probing_2023}. This measurement suggests that there is excess planetary flux at NIR wavelengths over what we would expect from thermal emission alone. This excess in planetary flux could be caused by the presence of aerosols over the dayside that reflect a certain fraction of the incoming stellar radiation. From the chemical equilibrium retrieval, we find that the cloud top pressure, atmospheric metallicity, and carbon-to-oxygen ratio are unconstrained and span the full range of their prior. Nevertheless, we observe that a certain region of the [M/H] and C/O parameter space appears to be excluded by the data, where scenarios of atmospheric metallicities above a hundred times solar with carbon-to-oxygen ratios between 0.6 and 0.8 are disfavored at more than 3\,$\sigma$ by the retrieval (see Figure \ref{fig:corner_chem_equi}). 

From the free chemistry retrieval, we find a brightness temperature of $T_\mathrm{B}=804_{-52}^{+46}$\,K and a reflected light geometric albedo of $A_\mathrm{g}=0.211_{-0.054}^{+0.050}$, which are consistent within 1\,$\sigma$ of the constraints from the chemical equilibrium retrieval.
We also find that the posterior probability of water, methane, carbon dioxide, carbon monoxide, and ammonia all span the full range of their prior (see Figure \ref{fig:corner_free_chem}), which is expected given the lack of any significant molecular features in the spectrum.
We use the brightness temperature and geometric albedo measurements from the chemical equilibrium retrieval, which are more conservative given their larger uncertainties, for the rest of our interpretation.

We further quantify the significance of the preference for a non-zero geometric albedo by comparing the maximum likelihood of retrievals with and without the geometric albedo. We measure an increase in the log-likelihood of 2.4 when including it, corresponding to a $p$-value of 0.09 (significance of 1.3\,$\sigma$). While this is not sufficient to claim that the inclusion of the geometric albedo in the retrievals is statistically preferred, we find that our measured dayside brightness temperature is in apparent conflict with previous studies when the geometric albedo is not considered. Indeed, without including the albedo, we measure a brightness temperature of $T_\mathrm{B} = 959_{-25}^{+21}$\,K--4.3 and 3.0\,$\sigma$ away from the brightness temperatures of $791\pm30$\,K and $871\pm16$\,K  measured from the Spitzer/IRAC 3.6- and 4.5-$\mu$m secondary eclipses in \citet{wong_hubble_2022}, respectively. This temperature is also in slight disagreement with \citet{bell_methane_2023}, who found the dayside temperature of the planet to be consistent with full redistribution ($T_\mathrm{eq}=825\pm90$\,K). When including the geometric albedo, our dayside brightness temperature of $811_{-70}^{+69}$\,K is in agreement with \citet{wong_hubble_2022} and \citet{bell_methane_2023}.

There is an apparent discrepancy between the significance of our measurement of short-wavelength flux at the light-curve fitting and atmospheric retrieval stages: we obtain a geometric albedo measurement that is more than 3\,$\sigma$ away from 0, but the white-light curve binned from the order 1 $<$1.5-$\mu$m data only produces a 1.2\,$\sigma$ measurement (see \S\ref{sec:eclipse_depths}). This inconsistency is most likely caused by the use of a GP in the light curve fit, which seemingly introduces a systematic error in the eclipse depth measurement that is independent of the actual light curve scatter, i.e., binning in wavelength before fitting does not improve the eclipse depth uncertainty.
When looking at the spectroscopic eclipse depth uncertainties, we observe that, in many spectral bins, we reach a precision that is equal to or better than that of the white-light curve fit. Moreover, if we bin all eclipse depth measurements below 1.5\,$\mu$m, the wavelengths from which the reflected light is most constrained, we obtain an eclipse depth of 33$\pm$11\,ppm. Although no single point within that wavelength range is detected at 3\,$\sigma$, they are all consistently slightly above 0, leading to the $\sim$3\,$\sigma$ measurement of a reflected light geometric albedo in the atmospheric retrieval.

\subsection{Interpretation of $A_g$ and $T_B$}
\label{sec:Ag_TB}

\begin{figure}
	\centering
	\includegraphics[width=\columnwidth]{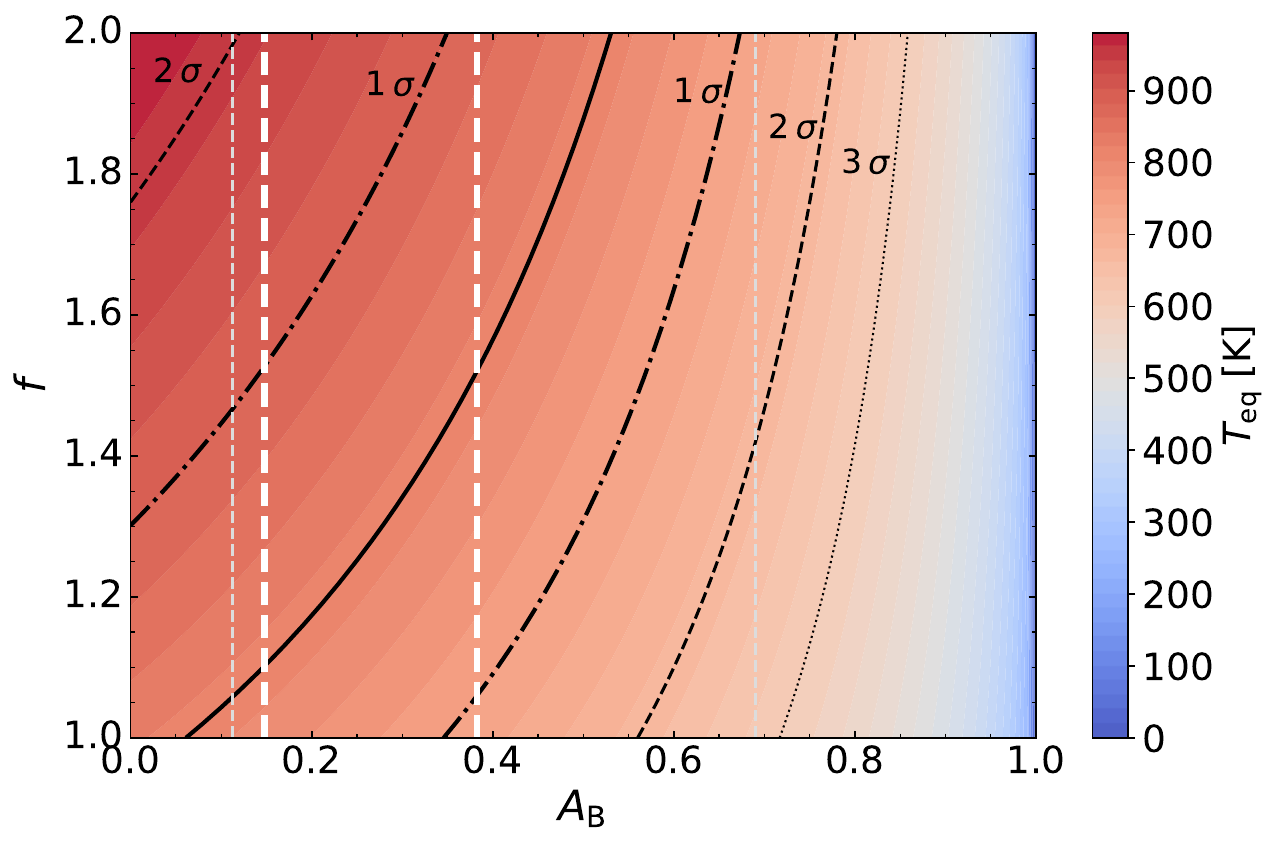}
    \caption{Equilibrium temperature as a function of the Bond albedo, $A_{\rm B}$, and the recirculation factor, $f$. Since WASP-80\,b is a warm gas giant, it is expected to distribute its energy at least uniformly across its dayside ($f=2$) or, at best, across its entire surface ($f=1$). The retrieved dayside brightness temperature is plotted in black. The median value is shown by a solid line, the 1-$\sigma$ confidence region is delimited by dash-dotted lines, the 2-$\sigma$ region by dashed lines, and the 3-$\sigma$ region by dotted lines. The gray dashed lines represent the technically possible 1-$\sigma$ limits on $A_\mathrm{B}$ ($0.113<A_\mathrm{B}<0.691$), while the white dashed lines show the more realistic 1-$\sigma$ limits on $A_{\rm B}$ ($0.148\lesssim A_\mathrm{B}\lesssim0.383$), which cover the possible values for $q$. Our measured dayside brightness temperature is consistent with full heat redistribution.
    \label{fig:T_f_AB}}
\end{figure}

With our measured brightness temperature and geometric albedo, we can estimate the values of the Bond albedo, $A_\mathrm{B}$, and heat recirculation factor, $f$. These quantities are related to the dayside equilibrium temperature via the following equation:

\begin{equation}
\label{eq:T_eq}
    T_{\mathrm{eq}}=T_{\mathrm{eff},*}\left(\frac{R_*}{2a}\right)^{1/2}[f(1-A_\mathrm{B})]^{1/4}\,,
\end{equation}

\noindent 
where $T_\mathrm{eff,*}$ is the effective temperature of the star \citep{charbonneau_detection_2005, rowe_upper_2006}.
Figure \ref{fig:T_f_AB} shows different equilibrium temperatures for WASP-80\,b for all possible values of Bond albedo and recirculation factor.
We can approximate the dayside equilibrium temperature by our measured brightness temperature, as we integrate the planetary emission over a relatively wide range of wavelengths, making our measured brightness temperature less sensitive to the presence or lack of atmospheric features.
In Figure \ref{fig:T_f_AB}, we have overplotted different confidence intervals for our measured dayside temperature
to visualize the possible values for the Bond albedo and redistribution factor. 

Moreover, we can relate the Bond albedo to the geometric albedo using the relation $A_\mathrm{B}=qA_\mathrm{g,TOT}$, where $A_\mathrm{g,TOT}$ is the geometric albedo integrated over all wavelengths and $q$ is the phase integral \citep{sudarsky_albedo_2000}. Given that WASP-80\,b is a warm gas giant, the value of the phase integral must be $1.0<q<1.5$ \citep{rowe_upper_2006, rowe_very_2008, schwartz_balancing_2015}. The lower limit is explained by the semi-infinite atmosphere limit for gas giants with thick atmospheres, while the upper limit is explained by the isotropic limit from a Lambert sphere \citep{sudarsky_albedo_2000, rowe_upper_2006, rowe_very_2008, schwartz_balancing_2015}. In fact, several gas giants have a higher Bond albedo than total geometric albedo, including hot Jupiters \citep{schwartz_balancing_2015, heng_closed-form_2021} and those in the Solar System \citep{rowe_upper_2006, li_less_2018}.

To constrain the Bond albedo, we use a similar formalism as in \cite{schwartz_balancing_2015} and constrain $A_\mathrm{g,TOT}$ using our measured reflected light geometric albedo for the SOSS bandpass. The lowest possible limit of $A_\mathrm{B}$ corresponds to the most extreme, albeit unlikely, scenario where the geometric albedo is 0 for all wavelengths outside the range of SOSS (and $q=1$). This limit is $A_\mathrm{B}>0.156^{+0.039}_{-0.043}$, given that the wavelength range of our SOSS spectrum contains 76\% of the stellar flux from WASP-80 (considering the same PHOENIX stellar model as in \S\ref{sec:Reduction}). On the opposite, we can put an absolute upper limit on $A_\mathrm{B}$ using Equation \ref{eq:T_eq} with our measured $T_\mathrm{B}$ and $f=2$, corresponding to inefficient recirculation of the heat. The upper limit on the Bond albedo in that case is $A_\mathrm{B}<0.531_{-0.162}^{+0.160}$. The 1-$\sigma$ lowest and highest technically possible limits on $A_\mathrm{B}$ are overplotted in gray in Figure \ref{fig:T_f_AB}.

We can place more realistic constraints on $A_\mathrm{B}$ by considering that our measured geometric albedo covers a significant portion of the wavelength range where reflected stellar light is strongest, since the peak of the spectral energy density function of WASP-80 coincides with the shortest wavelengths of our SOSS eclipse spectrum. Therefore, we expect the geometric albedo to be smaller at longer wavelengths than SOSS, but higher in the optical, where Rayleigh scattering and light scattering by clouds are more efficient \citep{marley_reflected_1999, demory_inference_2013, morley_thermal_2015}. We can therefore estimate $A_\mathrm{g,TOT}$ to be close to our measured $A_\mathrm{g}$ with SOSS, following the same gray scenario as described in \cite{schwartz_balancing_2015}. If we marginalize our constraint on $A_\mathrm{g}$ over the possible values of $q$, we obtain a 1-$\sigma$ confidence interval of $0.148\lesssim A_\mathrm{B}\lesssim0.383$ for the Bond albedo.
This interval is overplotted in white in Figure \ref{fig:T_f_AB} to better visualize the resulting possible values of the recirculation factor $f$. 

Theoretical models predict full heat redistribution for equilibrium temperatures and orbital period regimes similar to those of WASP-80\,b, as the propagation of atmospheric waves that can transport heat to the nightside is efficient \citep{showman_three-dimensional_2015, Komacek_2016, Komacek2017}. Full heat recirculation would also be consistent with the good agreement between the dayside and terminator spectra obtained by \cite{bell_methane_2023} with JWST/NIRCam. Figure \ref{fig:T_f_AB} shows that our possible range for $A_\mathrm{B}$, combined with our measured dayside brightness temperature, is indeed consistent with efficient heat redistribution ($f\sim1$). In contrast, the measured dayside temperature from the retrieval without the reflected light geometric albedo would lead to inefficient heat redistribution, reinforcing the need to include a geometric albedo in our analysis.

\begin{figure}
	\centering
        \includegraphics[width=\columnwidth]{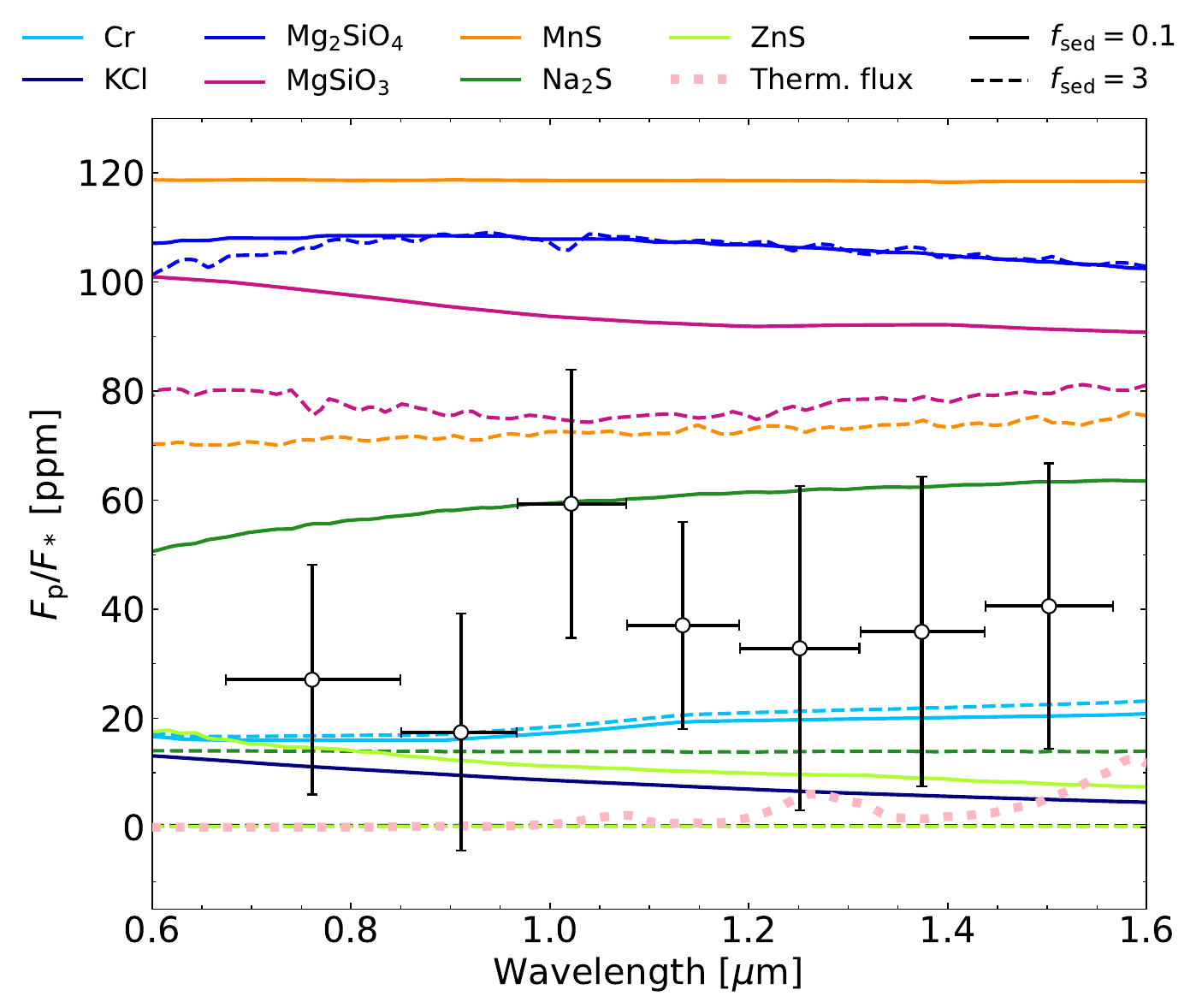}
    \caption{Reflected light models of WASP-80\,b for a variety of cloud species produced with \texttt{PICASO} and \texttt{VIRGA}, considering a Bond albedo of 0.25. The full and dashed lines correspond to models with sedimentation efficiencies $f_\mathrm{sed}$ of 0.1 and 3, respectively. The models are compared to our measured secondary eclipse spectrum (black and white points, binned by a factor of 2) for wavelengths below 1.6\,$\mu$m, where the contribution from thermal emission is minimal, as shown by the pink dotted line. The short-wavelength spectrum is best explained by either chromium (Cr), intermediate-sedimentation sodium sulfide (Na$_2$S), potassium chloride (KCl) or low-sedimentation zinc sulfide (ZnS) clouds.
    \label{fig:cloud_mods}}
\end{figure}

\begin{figure*}
	\centering
	\includegraphics[width=\textwidth]{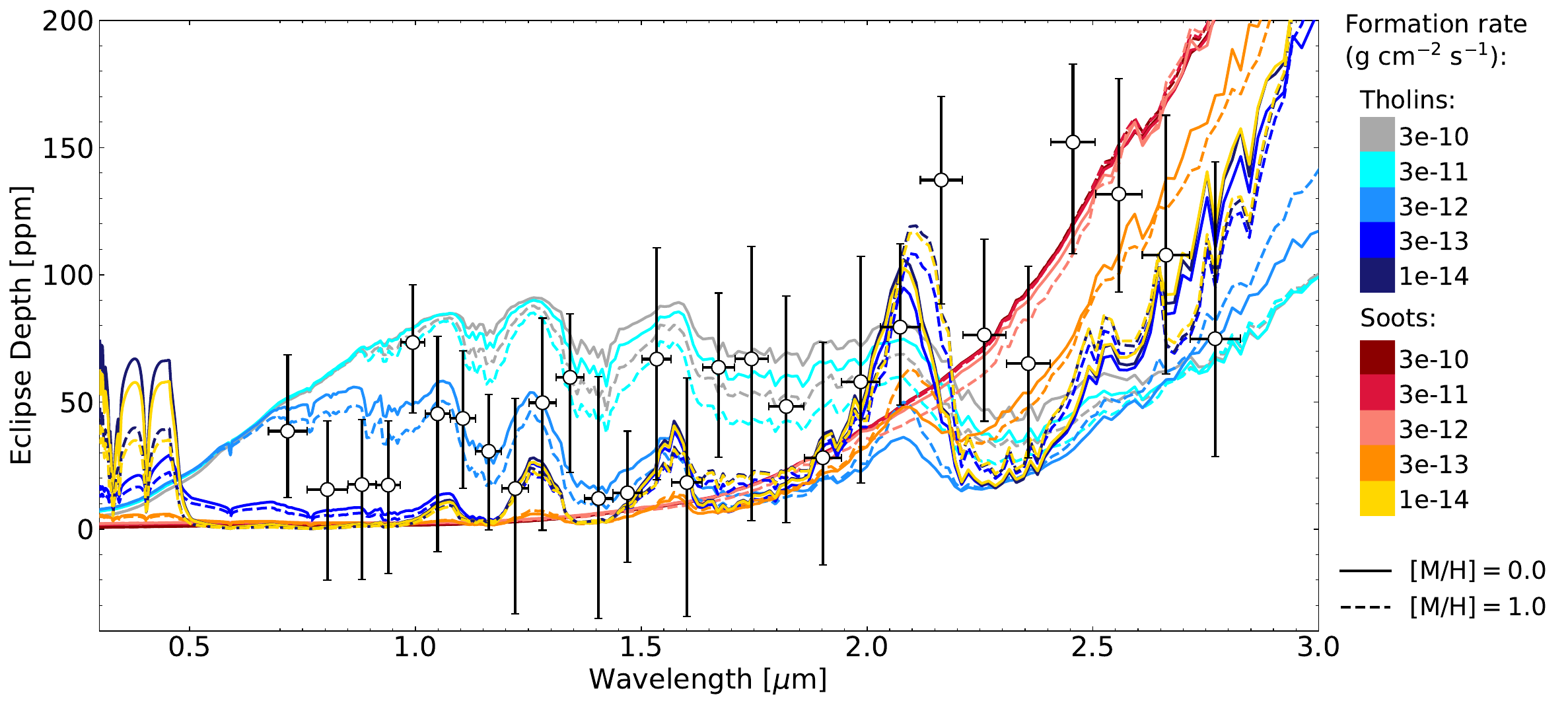}
    \caption{Haze (tholin and soot) models from \cite{jacobs_probing_2023} for two metallicities: $[\mathrm{M/H}]=0$ (solid) and $[\mathrm{M/H}]=1$ (dashed). The various formation rates are shown in cool colors for the tholins and in warm colors for the soots. The haze models are compared to the NIRISS/SOSS eclipse spectrum of WASP-80\,b. If the aerosols are hazes, they could consist of soots or tholins, with formation rates for tholins most likely lower than $10^{-11.5}$ g\,cm$^{-2}$\,s$^{-1}$.
    \label{fig:haze_models}}
\end{figure*}

\subsection{Clouds and hazes in the atmosphere of WASP-80\,b}
\label{sec:results_clouds}

With the aim to determine the type of aerosols that produce reflected light spectra consistent with our observations, we first compare our short-wavelength ($\lambda<$1.6\,$\mu$m) spectrum with our reflected light models produced with \texttt{PICASO}/\texttt{VIRGA} in Figure \ref{fig:cloud_mods}. We present only the models with $A_\mathrm{B}=0.25$, as this value is more likely to approximate the true Bond albedo of WASP-80\,b compared to $A_\mathrm{B}=0$. Additionally, the cloud models for both albedo values are very similar, as expected given their closely matching T-P profiles (see Figure \ref{fig:retrieval_constraints}).

We conduct a $\chi^2$ test between our short-wavelength spectrum and the cloud models and reject the species that are more than 3\,$\sigma$ away from the expected value of $\chi^2$, equal to the number of degrees of freedom ($\nu=14$\,data points$-1=13$), based on the following equation\footnote{From Section 7.2.1 of \cite{gregory_bayesian_2005}}:

\begin{equation}
\label{eq:n_sigma}
    n_\sigma = \frac{\chi^2 - \nu}{\sqrt{2\nu}}.
\end{equation}

\noindent The results for the reduced $\chi^2$ ($\chi^2_\nu$) and $n_\sigma$ are presented in Table \ref{tab:clouds}. Some values of $\chi^2_\nu$ are below unity (and their corresponding values of $n_\sigma$ negative) given the large error bars on our data.

The test allows us to rule out manganese sulfide (MnS) clouds, as well as forsterite (Mg$_2$SiO$_4$) and enstatite (MgSiO$_3$) clouds, as they all have $n_\sigma>3$\,$\sigma$ for both sedimentation efficiencies. The latter rejections were expected, as silicate species are predicted to dominate the aerosol opacity at equilibrium temperatures above 1500\,K--much higher than the temperature of WASP-80\,b--while transitioning to sulfide clouds and hazes at lower temperatures \citep{parmentier_transitions_2016, gao_aerosols_2021}. Silicate clouds would also be too reflective to match our observations.

Along with Figure \ref{fig:cloud_mods}, the $\chi^2$ test suggests that our short-wavelength spectrum is consistent with chromium (Cr[s]) or intermediate-sedimentation sodium sulfide (Na$_2$S) clouds. The reflectance spectrum of Cr[s] is relatively constant around 20\,ppm for both sedimentation efficiencies tested. As for the Na$_2$S clouds, their reflectance spectra show a stronger dependency with $f_\mathrm{sed}$ but could possibly explain our short-wavelength spectrum for an $f_\mathrm{sed}$ value that is between 0.1 and 3. This would be consistent with theoretical expectations, which predict the condensation of Na$_2$S in the temperature regime of WASP-80\,b \citep{marley_reflected_1999}. A combination of both species might also be possible, given that Cr[s] could provide the seed particles for sulfides like Na$_\mathrm{2}$S at temperatures similar to that of WASP-80\,b, as suggested by \cite{lee_dust_2018}. Based on the $\chi^2$ test, KCl and ZnS clouds are also consistent with our short-wavelength spectrum, although Figure \ref{fig:cloud_mods} shows that the latter is systematically above their spectra. Both species are expected to condense at the temperature of WASP-80\,b \citep{morley_neglected_2012, gao_aerosol_2020} and could also coexist, as KCl particles can act as condensation nuclei for ZnS \citep{gao_microphysics_2018, gao_aerosol_2020}.

We also compare our eclipse spectrum with haze models from \cite{jacobs_probing_2023}, as shown in Figure \ref{fig:haze_models}, for solar and $10\times$ solar atmosphere metallicities and column haze production rates ranging from $10^{-14}$ to $10^{-9.5}$ g\,cm$^{-2}$\,s$^{-1}$. Two haze compositions are considered: soots, which are made of complex hydrocarbons \citep{zahnle_thermometric_2009, morley_thermal_2015, lavvas_aerosol_2017, gao_aerosol_2020}, and tholins, which are created from organic compounds and nitrogen under strong UV radiation, like those in Titan's atmosphere \citep{khare_optical_1984, morley_thermal_2015}. Photochemistry is expected to be important on WASP-80\,b, as this planet receives high UV and X-ray irradiation from its host star \citep{king_xuv_2018}.

We conduct a similar $\chi^2$ test between our spectrum and the haze models, for which the results for the $\chi^2_\nu$ and $n_\sigma$ values are presented in Table \ref{tab:hazes} (here, $\nu=28$\,data points$-1=27$). The test indicates that if tholins are present in the dayside atmosphere of WASP-80\,b, their formation rate is most likely lower than $10^{-11.5}$\,g\,cm$^{-2}$\,s$^{-1}$, as the $n_\sigma$ values for higher rates are $>$3\,$\sigma$ and $\geq$2.45\,$\sigma$ for $\mathrm{[M/H]}=0.0$ and $\mathrm{[M/H]}=1.0$, respectively. This aligns with the conclusions drawn by \cite{jacobs_probing_2023}. If the aerosols are composed of soots, the $\chi^2$ test does not show any strong preference for a particular formation rate ($0.38\leq n_\sigma\leq1.17$). From Figure \ref{fig:haze_models}, the longest wavelengths ($>$2.7\,$\mu$m) suggest their production rate would also be low, with a value most likely no greater than $10^{-12.5}$\,g\,cm$^{-2}$\,s$^{-1}$, given the discrepancy between the expected eclipse depths and our measured spectrum at these wavelengths. Our eclipse depth measurements at short wavelengths ($<$1.5\,$\mu$m) also seem inconsistent with soot models with high production rate, as soots typically mute thermal emission features in the SOSS wavelength range rather than reflecting stellar light. 

\subsection{Comparison between NIRISS/SOSS and NIRCam}
\label{sec:NIRISS_vs_NIRCam}

As mentioned in \S\ref{sec:eclipse_depths}, we find a potential discrepancy between our measured eclipse depths with NIRISS/SOSS and those measured by \cite{bell_methane_2023} with NIRCam/F322W2 in the wavelength range that is common to both instruments. The disagreement between each bin ranges from 0.4 to 2.6\,$\sigma$, with a mean of 1.6\,$\sigma$, once the NIRCam data is binned down to the resolution of our NIRISS/SOSS spectrum. The disagreement is slightly less important with the additional reduction, as it ranges from 0.2 to 2.6\,$\sigma$, with a mean of 1.3\,$\sigma$.

This discrepancy could simply be the result of systematic noise, which is significant in the NIRISS/SOSS and NIRCam/F322W2 eclipse data. The wavelengths where the two spectra overlap correspond to regions of the NIRISS/SOSS and NIRCam detectors with lower throughput, where choices in data reduction are more likely to bias the secondary eclipse spectrum. Offsets between transmission spectra from different instruments or detectors have been observed before \citep{carter_benchmark_2024, wallack_jwst_2024}, and similar effects could also occur in eclipse spectroscopy, as predicted by \cite{barstow_transit_2015}. The difference between the two spectra could also be explained by different light curve modeling approaches, such as whether a GP model is included, as a GP can remove trends that might otherwise bias the retrieved eclipse depths. We also note that the NIRCam data points are not well represented by their best-fit emission spectrum model where the discrepancy is observed, as shown in Figure \ref{fig:spec_retrieval}. Their best-fit model for the emission spectrum has an average eclipse depth of $\sim$58 ppm in this wavelength range, supporting the expected eclipse detection in this region. However, our retrieved atmosphere model is consistent with theirs within its 2-$\sigma$ confidence region (see Figure \ref{fig:spec_retrieval}).

Another possible explanation could be temporal variability in the aerosol distribution in the atmosphere of WASP-80\,b. \cite{bell_methane_2023} did not include the effect of aerosols in their retrieval of their emission spectrum, but the NIRISS/SOSS data is consistent with the presence of reflective aerosols that increase the eclipse depths in the NIR. Given that the observations were made one year apart, temporal variability in the aerosol coverage could explain the difference in the measured eclipse depths. Such variability has been observed in brown dwarfs \citep{metchev_weather_2015, bowler_strong_2020, zhou_roaring_2022, vos_patchy_2023}, where their phase curve amplitudes vary over time on timescales of several days due to a heterogeneous cloud coverage caused by weather processes, such as radiative cloud feedback, cloud dissipation, storms, etc., as predicted by 3D circulation models \citep{tan_atmospheric_2019, cho_storms_2021, changeat_is_2024}. A similar argument was used to explain the temporal variations in the eclipse depths of the brown dwarf KELT-1\,b \citep{parviainen_temporal_2023}. It is unclear, however, if such processes also apply to colder gas giants ($T_\mathrm{eq}<2000$\,K).\\

\section{Conclusion}
\label{sec:Conclusion}

In this work, we presented the eclipse spectrum of WASP-80\,b obtained with JWST/NIRISS, providing the first eclipse measurements below 1.1\,$\mu$m for this planet, which enabled us to enhance our investigation of potential reflected light by its atmosphere. We find that the spectrum is most consistent with a combination of reflected light and thermal emission, as our measured dayside temperature from the retrieval that includes a reflected light geometric albedo is in better agreement with previous measurements of the temperature of WASP-80\,b. This result suggests the presence of aerosols over the dayside of this warm Jupiter.

We tightly constrain its reflected light geometric albedo for the first time between 0.68 and 2.83\,$\mu$m to a value of $A_\mathrm{g}=0.204_{-0.056}^{+0.051}$. From this measurement, we estimate the Bond albedo to be constrained within the interval $0.148\leq A_\mathrm{B}\leq0.383$ by marginalizing our constraint on $A_\mathrm{g}$ over the possible values of the phase integral. Given our measured dayside brightness temperature of $T_\mathrm{B}=811^{+70}_{-69}$\,K, our result for $A_\mathrm{B}$ is consistent with the prediction of effective heat recirculation from the dayside to the nightside of the planet.

By comparing cloud models with our short-wavelength ($\lambda<$1.6\,$\mu$m) spectrum, we rule out the possibility of manganese sulfide and silicate clouds in the observable atmosphere of WASP-80\,b, and find that Cr[s], Na$_\mathrm{2}$S, KCl or ZnS clouds can explain our observations. A comparison between our eclipse spectrum and haze models from \cite{jacobs_probing_2023} also shows that, if tholins are present in the dayside atmosphere of the planet, their production rate is likely smaller than $10^{-11.5}$\,g\,cm$^{-2}$\,s$^{-1}$. If soot hazes are present in its atmosphere, our results show no preference for a particular formation rate.

Although we are able to constrain the reflected light geometric albedo at more than 3\,$\sigma$ when it is included in the retrieval, the inclusion of this parameter in the retrieval is statistically preferred at less than 3\,$\sigma$. We therefore keep the possibility that no aerosols are present in the dayside atmosphere of this warm Jupiter.

Further work combining all known observations of WASP-80\,b, including data taken with JWST NIRCam (GTO 1185) and MIRI (GTO 1177), will shed further insights into the presence and composition of aerosols in its atmosphere.


\section*{Acknowledgments}

This work is based on observations made with the NASA/ESA/CSA James Webb Space Telescope. The data were obtained from the Mikulski Archive for Space Telescopes at the Space Telescope Science Institute, which is operated by the Association of Universities for Research in Astronomy, Inc., under NASA contract NAS 5-03127 for JWST. These observations are associated with program \#1201. K.M. acknowledges financial support from the Nature et technologies sector of the Fonds de Recherche du Québec (FRQ). The authors thank the Trottier Family Foundation for their support in the Trottier Institute for Research on Exoplanets (IREx) and the Trottier Space Institute (TSI). The authors thank IREx and the TSI for their financial support and dynamic intellectual environment. The authors are grateful for financial support from the FRQ and NSERC. NBC acknowledges support from an NSERC Discovery Grant, a Tier 2 Canada Research Chair, and an Arthur B.\ McDonald Fellowship. D.J.\ is supported by NRC Canada and by an NSERC Discovery Grant. This project was undertaken with the financial support of the Canadian Space Agency. S.P. acknowledges the financial support of the Swiss National Science Foundation (SNSF) under grant 51NF40\_205606. This project has been carried out within the framework of the National Centre of Competence in Research PlanetS supported by the SNSF. R.A. acknowledges the SNSF support under the Post-Doc Mobility grant P500PT\_222212. L.D.\ acknowledges support from the Banting Postdoctoral Fellowship program, administered by the Government of Canada. C.P.-G. acknowledges support from the NSERC Vanier scholarship.  J.F.R. acknowledges support from an NSERC Discovery Grant, a Tier 2 Canada Research Chair and computational resources from Digital Alliance Canada. The authors would also like to thank the Centre de recherche en astrophysique du Québec. The authors would like to thank Taylor J. Bell and Thomas P. Greene for useful and insightful discussions. The authors would also like to thank Bob Jacobs for sharing some of his models to help improve the analysis. This research has made use of the NASA Exoplanet Archive, which is operated by the California Institute of Technology, under contract with the National Aeronautics and Space Administration under the Exoplanet Exploration Program. This research was enabled in part by support provided by Calcul Québec (\url{www.calculquebec.ca/}) and the Digital Research Alliance of Canada (\url{https://alliancecan.ca/}). The authors thank the anonymous referee for their careful review and insightful comments, which improved the quality of this work.


\facilities{JWST, Exoplanet Archive, MAST}

\software{
\texttt{astropy} \citep{the_astropy_collaboration_astropy_2013, the_astropy_collaboration_astropy_2018},
\texttt{batman} \citep{kreidberg_batman_2015},
\texttt{celerite} \citep{foreman-mackey_fast_2017},
\texttt{dynesty} \citep{speagle_dynesty_2020, koposov_joshspeagledynesty_2023}
\texttt{jwst} \citep{bushouse_jwst_2022},
\texttt{matplotlib} \citep{hunter_matplotlib_2007},
\texttt{numpy} \citep{harris_array_2020},
\texttt{scipy} \citep{virtanen_scipy_2020},
\texttt{exoTEDRF} \citep{radica_awesome_2023, Radica2024},
\texttt{SCARLET} \citep{benneke_atmospheric_2012},
\texttt{PICASO} \citep{batalha_exoplanet_2019},
\texttt{VIRGA} \citep{Rooney2022}
}

\bibliography{ref}
\bibliographystyle{aasjournal}

\appendix
\renewcommand{\thefigure}{A\arabic{figure}} 
\setcounter{figure}{0} 

\renewcommand{\thetable}{A\arabic{table}} 
\setcounter{table}{0} 

\section{Additional Reduction}\label{sec:add_reduc}

\begin{figure*}[h!]
	\centering
	\includegraphics[width=0.75\textwidth]{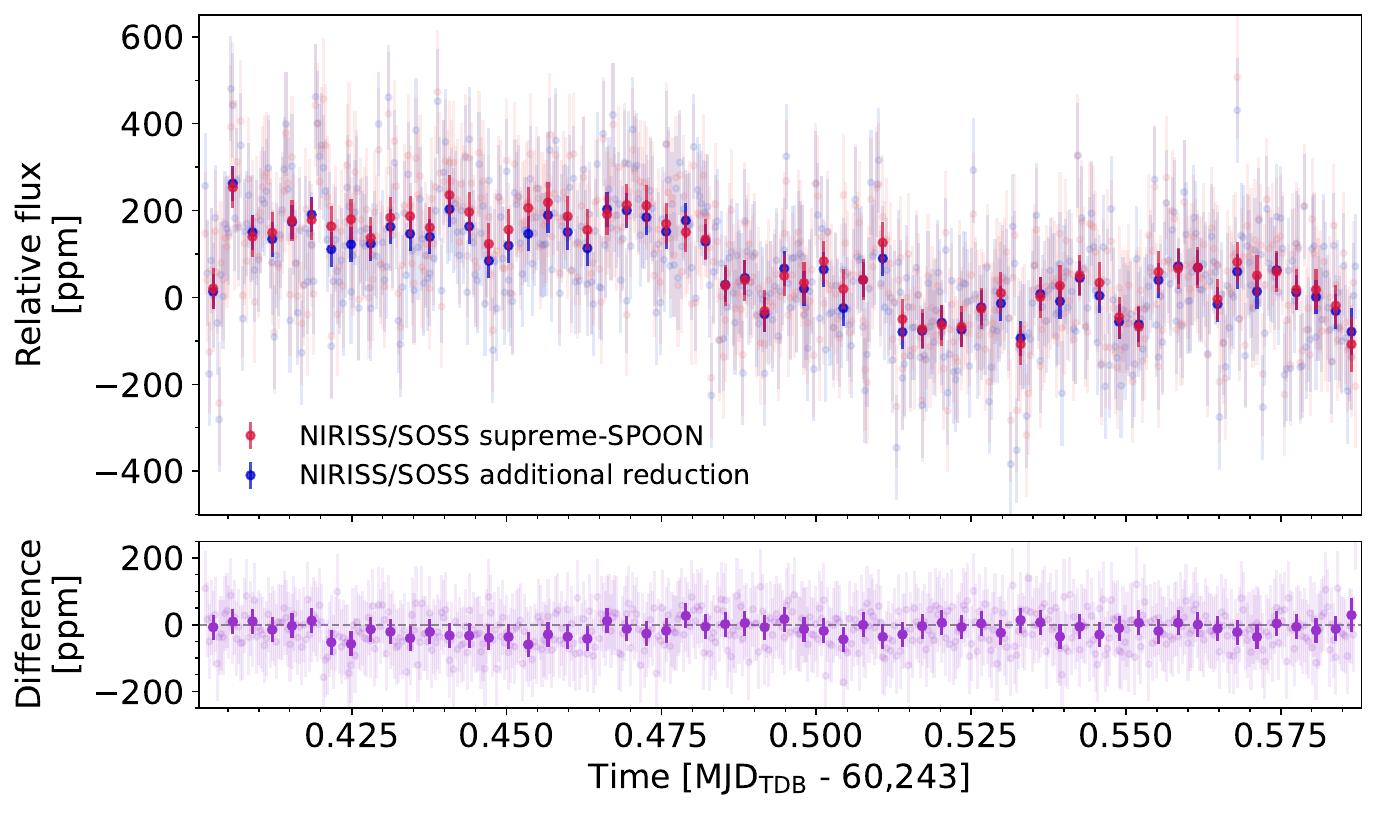}
    \caption{\textbf{Top:} Raw NIRISS/SOSS order 1 (0.85-2.83\,$\mu$m) white-light curves with 1-$\sigma$ error bars for the supreme-SPOON reduction (red) and the additional one (blue) with their temporally binned points. \textbf{Bottom:} Difference between the white-light curve from the additional reduction and the one from \texttt{supreme-SPOON}. The error bars on the residuals represent the maximum ones between the two reductions. In general, both white-light curves follow the same trends.
    \label{fig:2wlcs}}
\end{figure*}

A custom, fully independent pipeline was developed to validate the analysis of SOSS relative to other reductions and techniques. The pipeline applies instrumental corrections to imaging data, extracts spectrophotometry and corrects for aperture contamination from spectral overlap that is common for SOSS observations. Analysis uses Stage 0 data products (e.g., \texttt{uncal.fits}) and applies JWST reference files. The pipeline then uses a two-step process (described below) to apply a 1/$f$ noise group-level correction using difference image techniques. The premise is that difference imaging will provide differential spectrophotometry measurements, which are then scaled by analyzing a median deep-stack image. This approach should be robust against chromatic aperture dilution from other stellar sources in the field. Figure \ref{fig:spectra} shows a comparison of the extracted eclipse spectrum of WASP-80\,b against supreme-SPOON (see \S\ref{sec:Reduction}). A fit to the white-light curve using the method described in \S\ref{sec:wlc_fit} estimates the white-noise component to be $112.8_{-3.6}^{+4.7}$\,ppm for the custom reduction and $106.0_{-2.9}^{+5.0}$\,ppm for supreme-SPOON, with small differences likely due to the aggressiveness of correcting for cosmic-ray hits and hot-pixels. Overall, the agreement between the two pipelines provides strong evidence that any systematics present in the extracted spectrophotometric light curve, as shown in Figure \ref{fig:2wlcs}, are unlikely due to data processing errors.

Non-linearity and superbias corrections were made using the same reference files as adopted in \S\ref{sec:Reduction}. Reference pixels corrections used simple means with 3\,$\sigma$ rejection for outliers. The pipeline uses a saturation reference map to flag any group during an exposure that exceeds saturation. Saturated group values for each pixel were subsequently excluded when integrating ``up-the-ramp" to estimate the electron counts for each pixel for each integration. Bad pixels were flagged as due to either total saturation or non-zero values of bit 0 from the data-quality flags, and replaced by the mean of a 3$\times$3 box centered on the offending group value.

A two-step iterative process was used to calculate count rates and to correct for 1/$f$ noise using difference images. Figure \ref{fig:Reduction_Steps} shows the first, second and third order spectrum for WASP-80 as well as a field star. Difference imaging reduces the impact of the PSF wings on any calculation that depends on a statistical average to estimate column averages of the background level, which we interrupt as an estimate to correct for 1/$f$ noise. The first stage used linear-regression of ramp values to estimate residual bias ($\it zpt$) and the integrated electron count rate. Column and row means of the $\it zpt$ values for each integration were computed and then used as an additional bias correction for each group image. Differences relative to the superbias were found at the ADU level. A stacked median image for each group step was calculated using all integrations. This produced four 256$\times$2048 deep-stacks based on 668 integrations. Thus, there were four samples of the ramp for each integration. Difference images for each group set were then used to measure the 1/$f$ noise correct using column medians. Only pixels more than 9 pixels away from the spectral trace were used when calculating the median. The second stage applied 1/$f$ corrections at the group level and we repeated the ramp integration to calculate count rates based on a robust mean of forward-differences, with outliers excluded at the 2-$\sigma$ level. 

A median image was then used to calculate difference time-series images by subtracting the median image from each ramp-corrected integration. Aperture photometry along the spectral trace for the first and second order used an aperture radius of 16 pixels to extract the differential spectrum.  

\begin{figure*}[h]
	\centering
	\includegraphics[width=\textwidth]{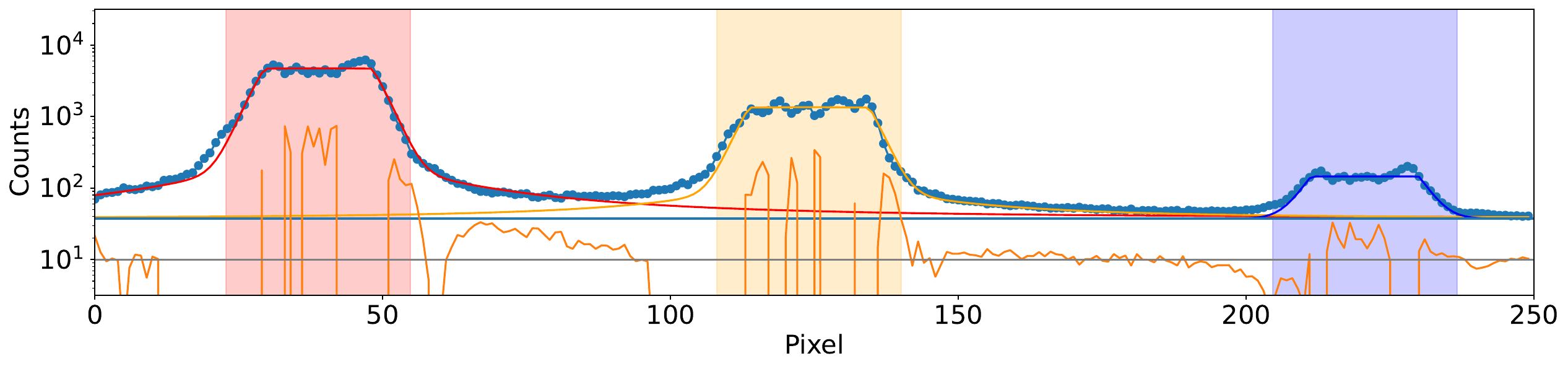}
    \includegraphics[width=\textwidth]{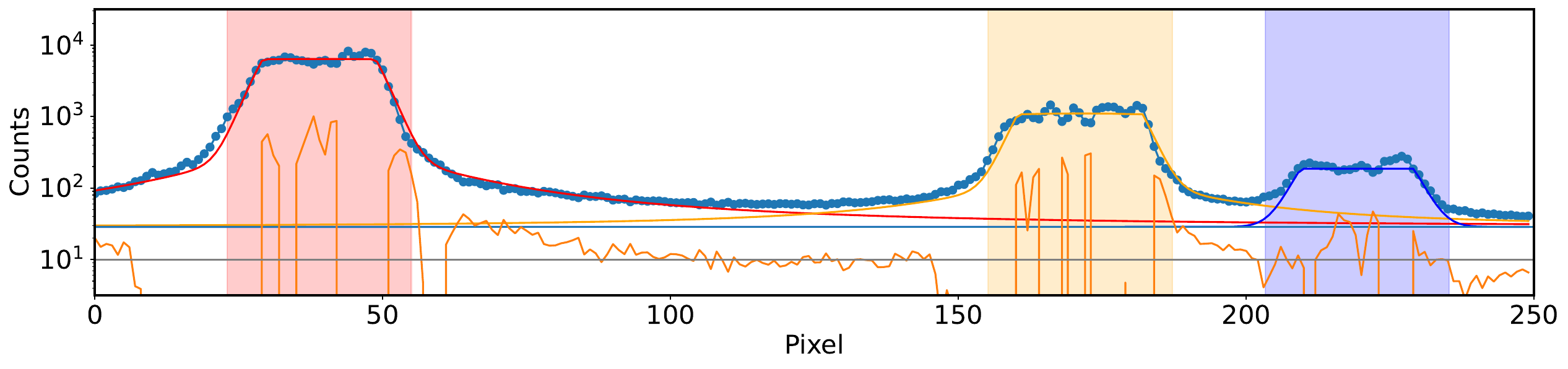}
    \caption{\textbf{Top:} A cross-section(column 1300) of the SOSS 2D image of the WASP-80 field. The photometric aperture used to extract the spectrum in the additional reduction of the first and second-order traces are highlighted by the light-red and orange boxes.
    The imaged field star is highlighted by the light-blue box.
    The red, orange and blue lines show the PSF model used to model distribution of starlight across the field. The PSF model was used to estimate photometric contamination inside the aperture of each trace and repeated for each column to construct a decontamination model to accurately measure eclipse depths in the additional reduction. \textbf{Bottom:} Same as top panel, but for column 1575, which is the last column used in the extraction of order 2. The contamination from the imaged field star on the order 2 extracted flux is $<1\%$ at this column.
    \label{fig:psf_contam_model}}
\end{figure*}

We estimated the median stellar flux using the deep-stack median image. This allowed the results from difference imaging to be converted to relative photometry. As observed in Figure \ref{fig:Reduction_Steps}, there is physical overlap between the first and second-order traces of WASP-80. Additionally, the extended PSF wings from the first-order spectrum will contaminate the second order spectrum. We estimated the contamination for each order by fitting a {\it capped}-Lorentzian model to each column centered on the position of each visible trace on the spectrum. Two examples of the fitted model are shown in Figure \ref{fig:psf_contam_model}. The best fit model for each column was then integrated over the photometric aperture to calculate the amount of dilution present. The model also corrected for background due to scattered zodiacal light. We found that contamination of the first order trace peaks at the red-edge of the spectrum near 2.8\,$\mu$m with a total contribution of 200\,ADU from the second order trace corresponding to a contamination of less than 2\%. Additionally, PSF blending from the second order near 1.5\,$\mu$m contributes contamination less than 0.15\%. The second order trace has significantly more contamination peaking at 80\% near 1.25\,$\mu$m and a complete loss of the trace for wavelengths shorter than $\sim$0.67\,$\mu$m due to the field star. We extracted aperture spectrophotometry from the stacked median image and applied corrections for contamination and background based on our PSF model. The photometry from the median image was then used to convert the difference imaging photometry to relative photometry as presented in Figure \ref{fig:spectra}.\\

\section{Additional Figures}

\begin{figure*}[h]
	\centering
    \includegraphics[width=1.0\textwidth]{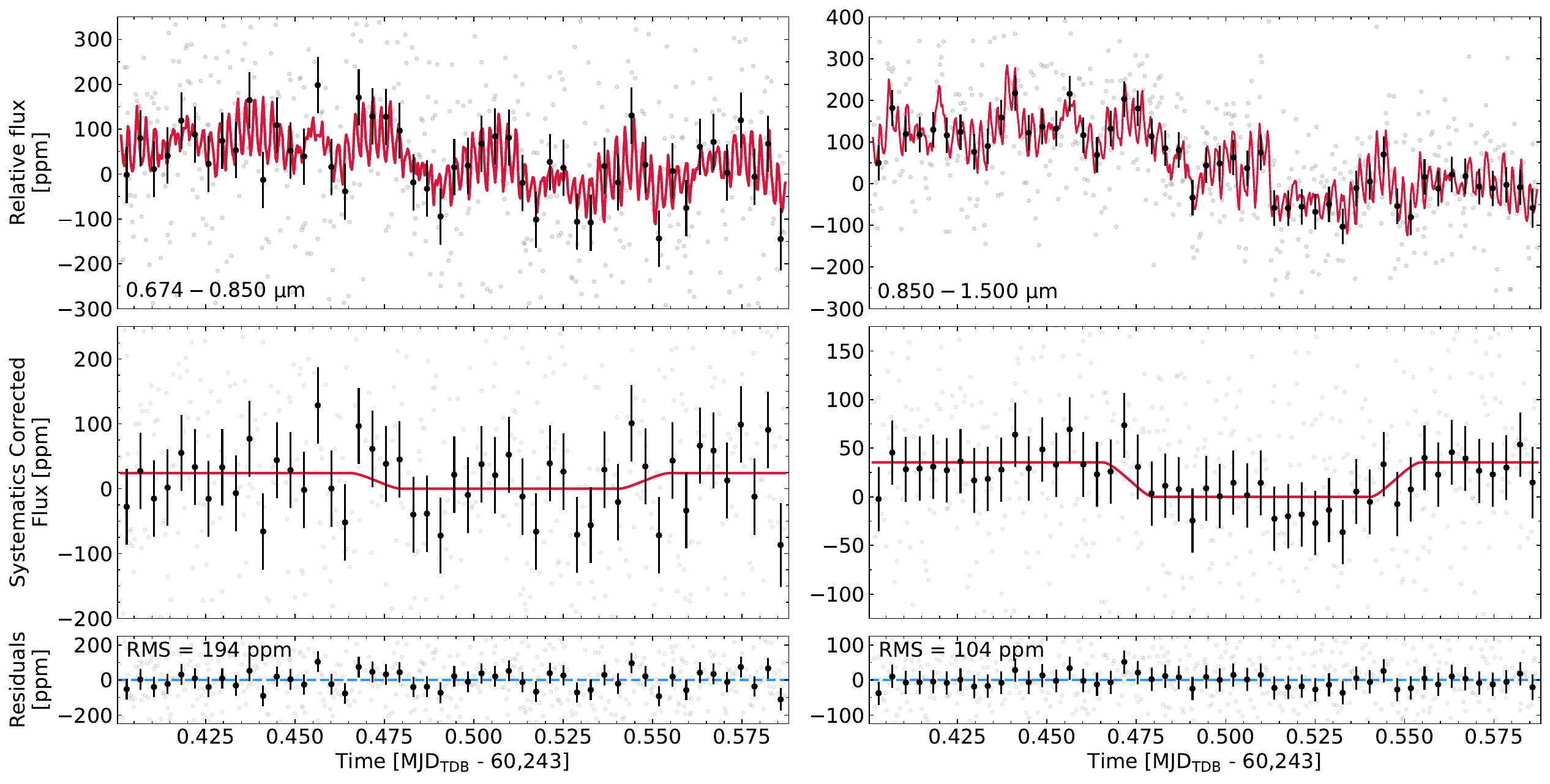}
    \caption{\textbf{Left:} Order 2 white-light curve fit (0.674 -- 0.850\,$\mu$m). \textbf{Right:} Light curve fit for a large bin of the order 1 trace spanning 0.850 -- 1.500\,$\mu$m. Both light curves show similar systematics, supporting the need to model the order 2 light curves like the order 1 light curves.
    \textbf{Top:} Raw NIRISS/SOSS light curves (gray) and their temporally binned points with 1-\,$\sigma$ error bars (black). The best-fit models, including the eclipse and the systematics, are overlapped in red. The retrieved eclipse depth for the order 2 white-light curve (left) is $26_{-37}^{+27}$\,ppm and the one for the short-wavelength order 1 bin (right) is $34_{-28}^{+25}$\,ppm. The latter measurement corresponds to a tentative detection of reflected light at short wavelengths.
    \textbf{Middle:} Detrended light curves (gray, binned points in black) once systematics are removed, along with the eclipse models (red). In each case, the error bars on the binned points are computed from the best-fit scatter parameter on the unbinned points (left: $\sigma$ = 194.9 ppm, right: $\sigma$ = 110.0 ppm). \textbf{Bottom:} Residuals to the best-fit models and their RMS values (based on the gray unbinned points).
    \label{fig:o1below15}}
\end{figure*}

\begin{figure*}[h!]
	\centering
	\includegraphics[width=0.9\textwidth]{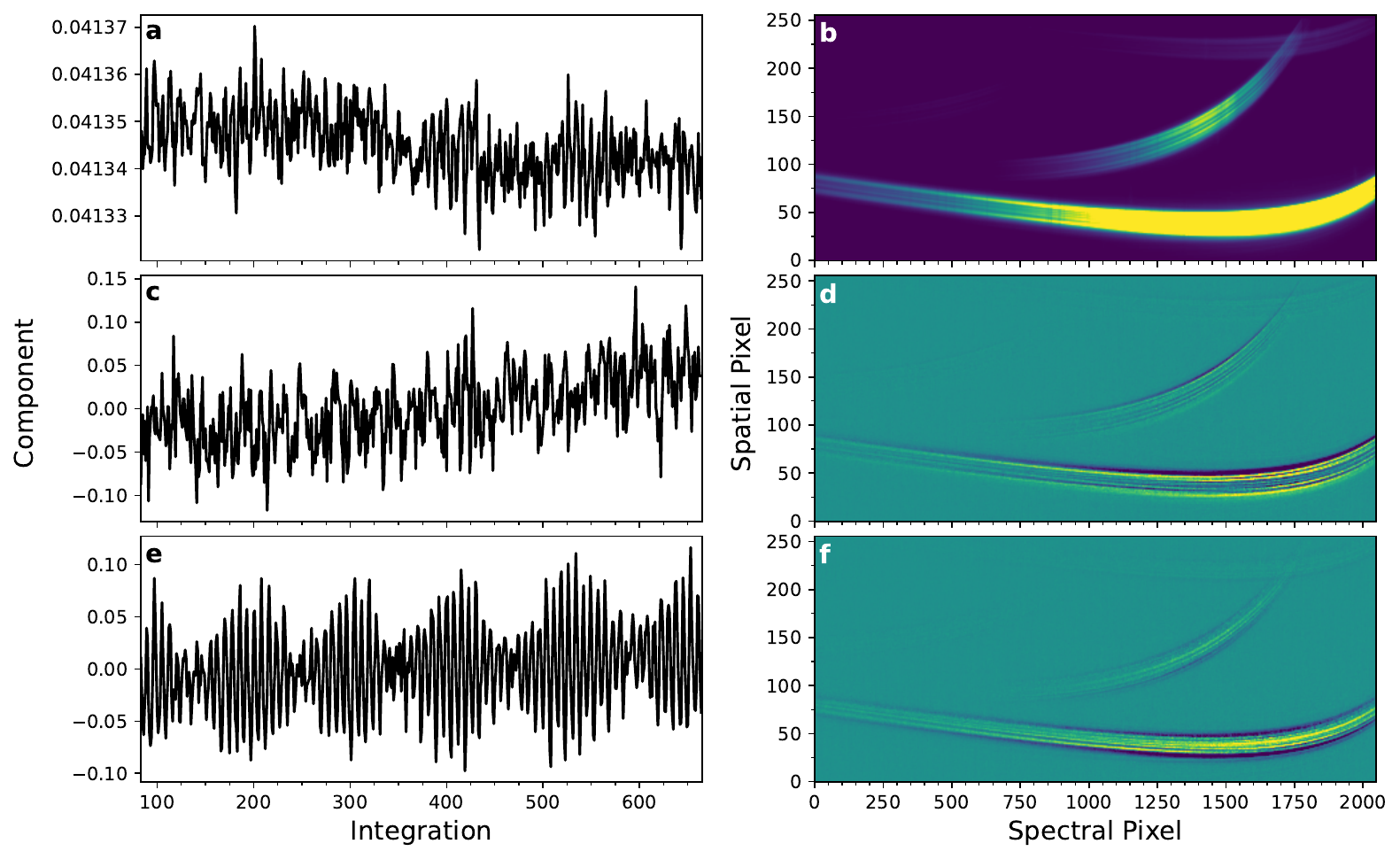}
    \caption{Results from the principal component analysis of the calibrated data products (without considering the integrations before the tilt event).
    \textbf{(a)} Eigenvalue of the first principal component, which clearly contains the eclipse.
    \textbf{(b)} First principal component, i.e., the one with the highest variance in the time series.
    \textbf{(c)} Eigenvalue of the second principal component.
    \textbf{(d)} Second principal component. This component represents changes in the spatial direction of the traces, resulting in trade of flux between the upper and lower parts of the traces.
    \textbf{(e)} Eigenvalue of the third principal component. This beating pattern is caused by temperature variations in the instrument electronics compartment. We use this component to detrend our light curves.
    \textbf{(f)} Third principal component. This component results in changes in the FWHM of the traces, as shown by the trade of flux between the wings and center of the traces.
    \label{fig:PCA}}
\end{figure*}

\begin{figure*}[h]
	\centering
	\includegraphics[width=0.6\textwidth]{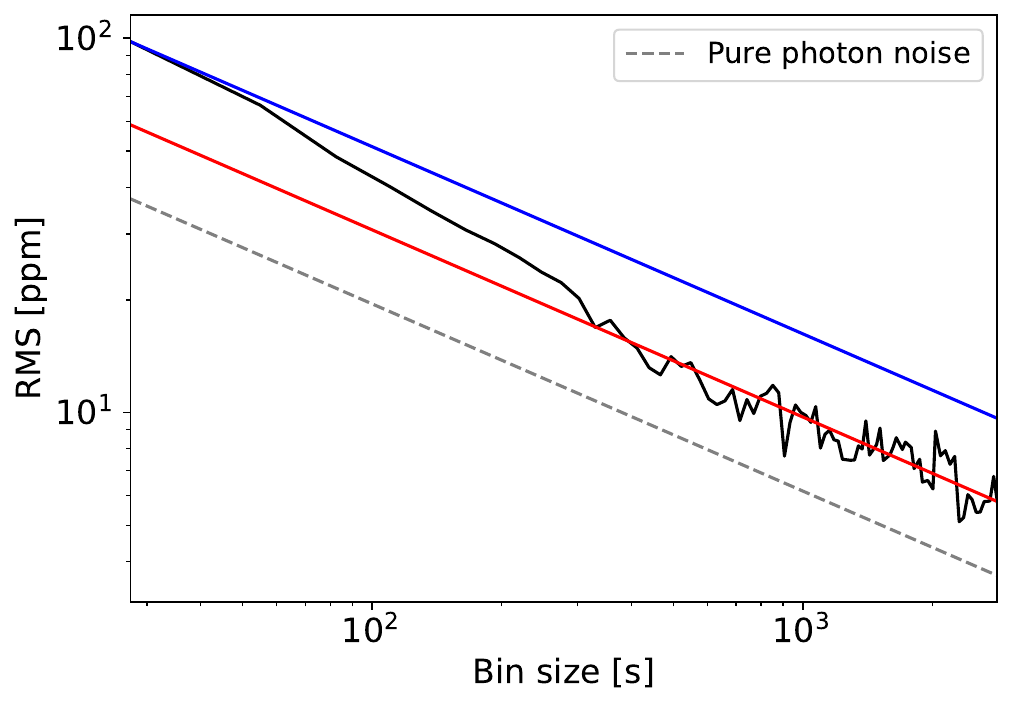}
    \caption{Root mean square (RMS) of the residuals from the order 1 white-light curve fit (black line) as a function of bin size. The blue line shows the decrease of the RMS for Poisson noise. Our residuals fall below this line but they follow the Poisson law for bin sizes larger than $\sim300$ s, as shown by the red line. This can be explained by the presence of remaining correlated noise at bin sizes smaller than the frequencies that the GP removes. Pure photon noise is represented by the gray dashed line. Our noise level is $\sim$1.5$\times$ the photon noise.
    \label{fig:allan}}
\end{figure*}

\begin{figure*}[h]
	\centering
	\includegraphics[width=0.9\textwidth]{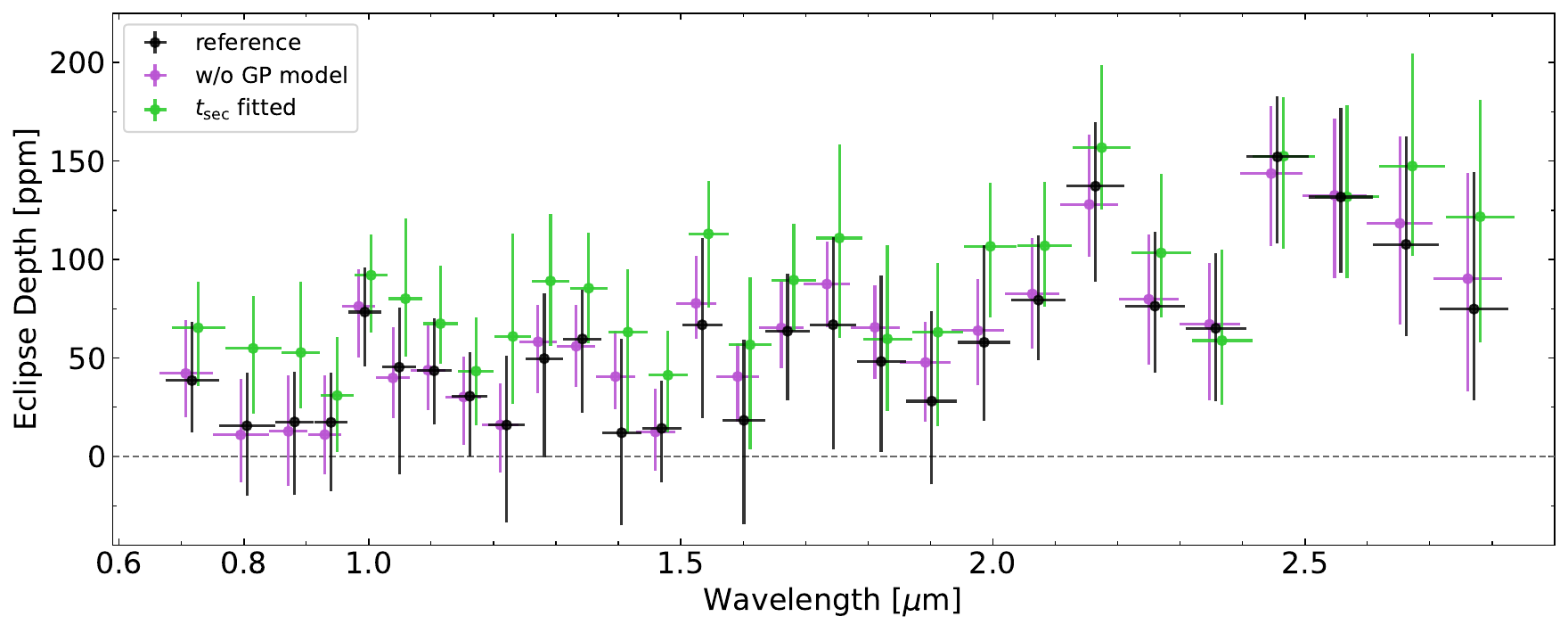}
    \caption{NIRISS/SOSS eclipse spectrum of WASP-80\,b according to different modeling approaches. Each eclipse depth is shown with its 1-$\sigma$ error bars. The black points represent the reference spectrum used for the analysis. The purple points, shifted by -0.01 $\mu$m for clarity, show the spectrum obtained without using a GP model in the light curve fitting. The inclusion of a GP in the light curve fit produces more realistic error bars on the spectrum. The green spectrum, shifted by +0.01 $\mu$m, follows the same modeling approach as the reference spectrum, but the time of mid-eclipse was fixed to the best-fit value from the white-light curve fit in which it was allowed to vary. This time is delayed by 12.8 minutes compared to the expected value.
    \label{fig:effects_model_spectra}}
\end{figure*}

\begin{figure*}[h]
	\centering
	\includegraphics[width=0.8\textwidth]{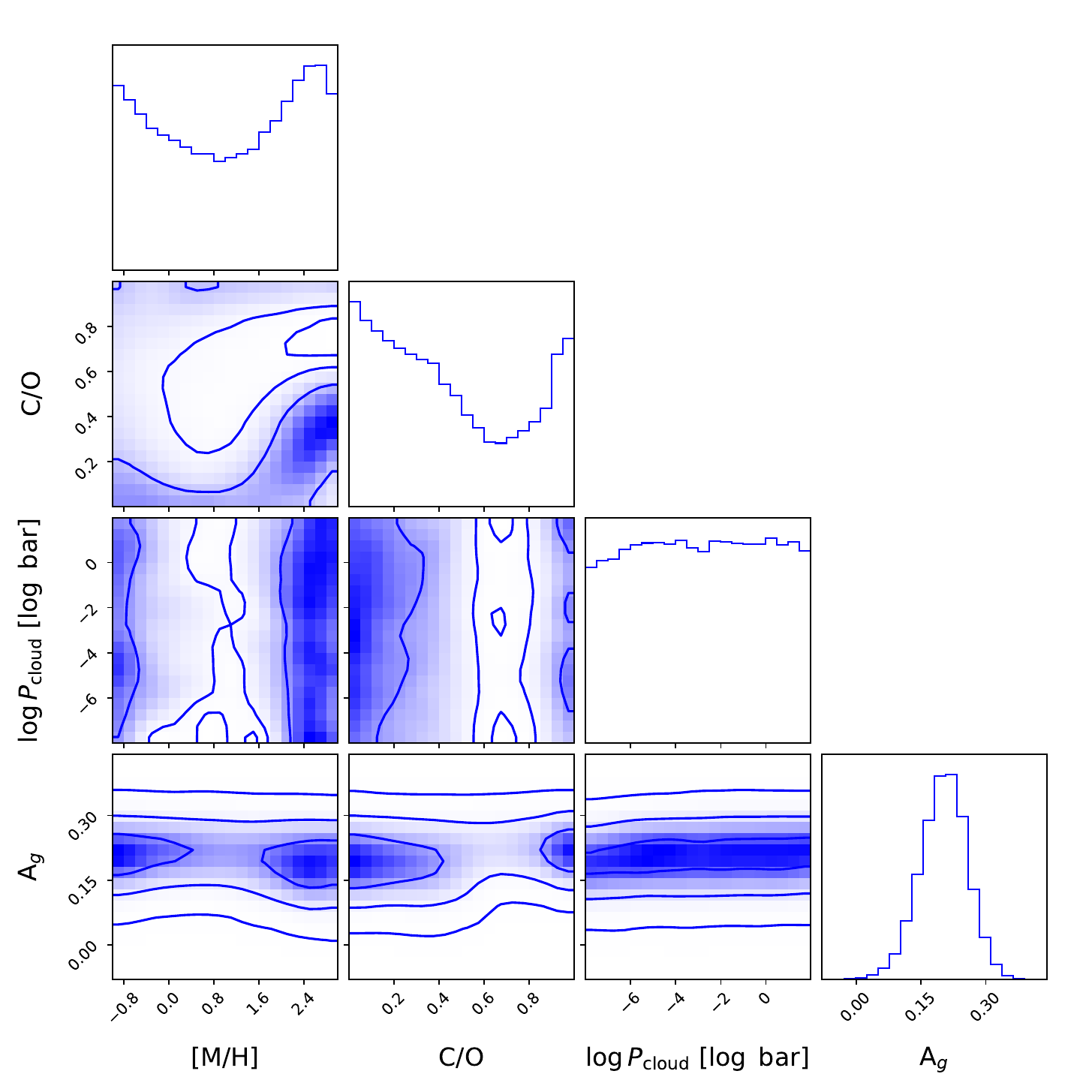}
    \caption{Corner plot of the parameters included in the chemical equilibrium retrieval, except for the 10 temperature-pressure profile parameters.
    \label{fig:corner_chem_equi}}
\end{figure*}

\begin{figure*}[h]
	\centering
	\includegraphics[width=\textwidth]{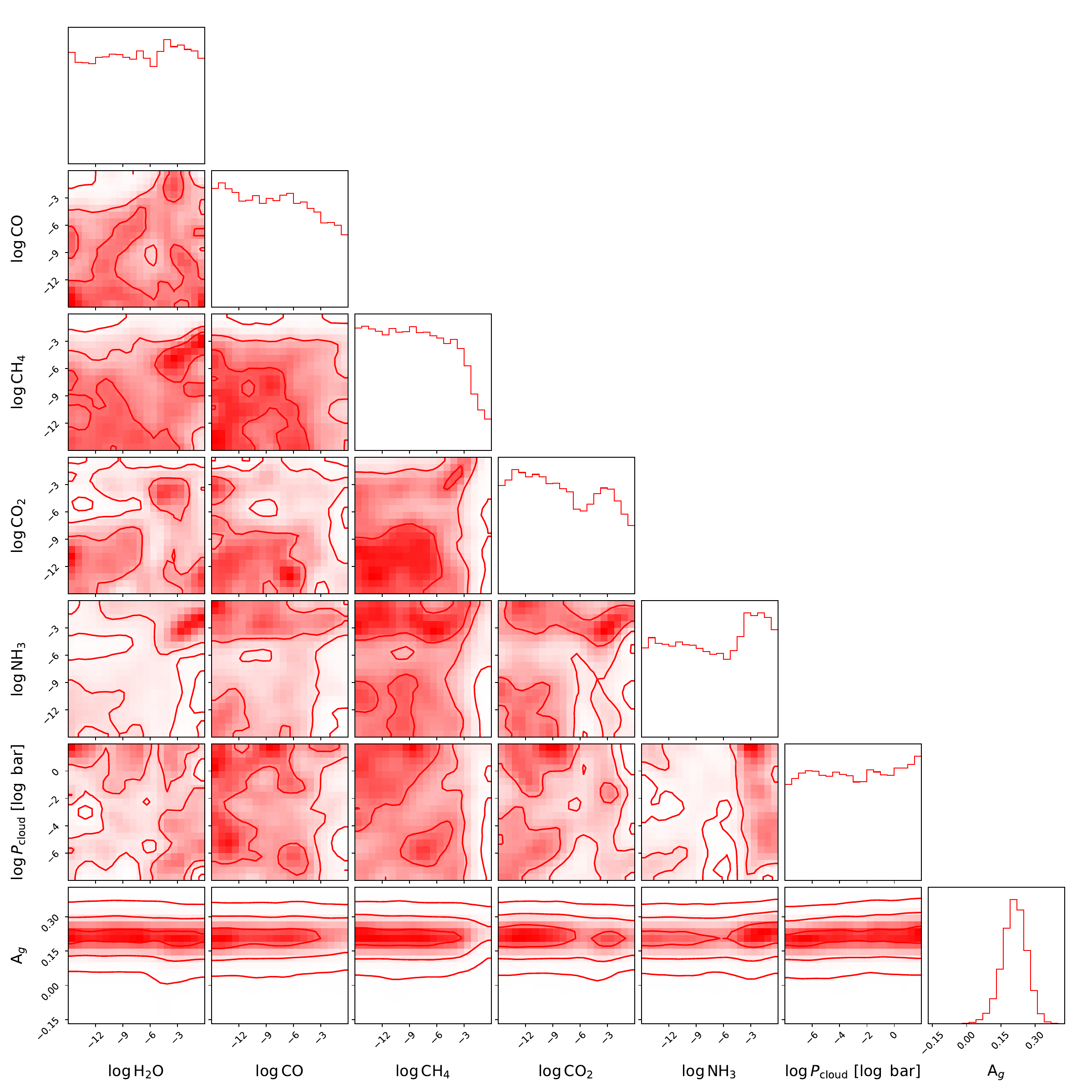}
    \caption{Corner plot of the parameters included in the free chemistry retrieval, except for the 10 temperature-pressure profile parameters.
    \label{fig:corner_free_chem}}
\end{figure*}

\clearpage

\section{Additional Tables}

\begin{table}[h!]
\begin{center}
\caption{Priors and results for the order 1 white-light curve fit for the \texttt{supreme-SPOON} and additional reductions. For each reduction, the results include the values of each parameter for the model with highest likelihood and the highest posterior density (HPD) interval of each parameter.}
\label{tab:4} 
\begin{tabular}{|lc||cc||cc|}\hline
& & \multicolumn{2}{c||}{\texttt{supreme-SPOON}} & \multicolumn{2}{c|}{Additional Reduction} \\
\hline
\multirow{ 2}{*}{\textbf{Parameters}} & \multirow{ 2}{*}{\textbf{Priors}} & \textbf{Highest} & \textbf{HPD} & \textbf{Highest} & \textbf{HPD} \\
&  & \textbf{Likelihood} & \textbf{Interval} & \textbf{Likelihood} & \textbf{Interval} \\
\hline
\hline
$a/R_*$ & $\mathcal{N}[12.643, 0.032]$ & 12.646 & $12.645_{-0.031}^{+0.033}$ &
12.645 & $12.643_{-0.030}^{+0.033}$\\
$i$ (deg) & $\mathcal{N}[88.938, 0.059]$ & 88.868 & $88.934_{-0.054}^{+0.064}$ &
89.036 & $88.937_{-0.062}^{+0.057}$\\
$F_{\rm p}/F_*$ (ppm) & $\mathcal{U}[-500,500]$ & 69 & $65_{-35}^{+28}$ & 
62 & $54_{-24}^{+30}$\\
$\sigma$ (ppm) & $\mathcal{U}[0,500]$ & 102.9 & $106.0_{-2.9}^{+5.0}$ & 
112.5 & $112.8_{-3.6}^{+4.7}$ \\
$d$ (ppm) & $\mathcal{U}[-250,250]$ & 36 & $44_{-23}^{+22}$ & 
28 & $31_{-19}^{+21}$\\
$b$ & $\mathcal{U}[-1,1]$ & $-1.06\times 10^{-3}$ & $-1.06_{-0.29}^{+0.21}\times 10^{-3}$ & $-1.00\times 10^{-3}$ & $-1.03_{-0.22}^{+0.24}\times 10^{-3}$\\
$c$ & $\mathcal{U}[-1,1]$ & $2.58\times 10^{-5}$ & $2.76_{-0.48}^{+0.50}\times 10^{-5}$ & $2.32\times 10^{-5}$ & $2.26_{-0.51}^{+0.54}\times 10^{-5}$\\
$a_{\rm GP}$ (ppm) & $\mathcal{LU}[1,200]$ & 57 & $55_{-10}^{+11}$ & 
51 & $52_{-7}^{+12}$\\
$\rho_{\rm GP}$ (min) & $\mathcal{LU}[1,60]$ & 2.5 & $3.1_{-1.6}^{+5.2}$ & 
2.5 & $3.0_{-1.5}^{+3.7}$\\
\hline
\end{tabular}
\end{center}
\end{table}

\begin{table}[h!]
\begin{center}
\caption{$\chi_\nu^2$ values between the short-wavelength ($<$$1.6$\,$\mu$m) NIRISS/SOSS eclipse spectrum of WASP-80\,b and the cloud models with the respective $n_\sigma$ values computed from Equation \ref{eq:n_sigma} considering 13 degrees of freedom. The sedimentation efficiency is denoted by $f_{\mathrm{sed}}$.}
\label{tab:clouds}
\begin{tabular}{|l||c|c||c|c||c|c||c|c||c|c||c|c||c|c|}
\hline
 & \multicolumn{2}{c||}{Cr} & \multicolumn{2}{c||}{Na$_2$S} & \multicolumn{2}{c||}{KCl} & \multicolumn{2}{c||}{ZnS} & \multicolumn{2}{c||}{MnS} & \multicolumn{2}{c||}{MgSiO$_3$} & \multicolumn{2}{c|}{Mg$_2$SiO$_4$} \\
\hline
\hline
$f_{\mathrm{sed}}$ & $\chi_\nu^2$ & $n_\sigma$ & $\chi_\nu^2$ & $n_\sigma$ & $\chi_\nu^2$ & $n_\sigma$ & $\chi_\nu^2$ & $n_\sigma$ & $\chi_\nu^2$ & $n_\sigma$ & $\chi_\nu^2$ & $n_\sigma$ & $\chi_\nu^2$ & $n_\sigma$ \\
\hline
0.1  & 0.66 & -0.86 & 1.34 & 0.86 & 1.08 & 0.21 & 0.94 & -0.14 & 9.96 & 22.86 & 5.19 & 10.69 & 7.51 & 16.61 \\
\hline
3 & 0.63 & -0.95 & 0.82 & -0.47 & 1.53 & 1.34 & 1.53 & 1.36 & 2.40 & 3.57 & 2.88 & 4.80 & 7.48 & 16.53 \\
\hline
\end{tabular}
\end{center}
\end{table}

\begin{table}[h!]
\begin{center}
\caption{$\chi^2_\nu$ values between the NIRISS/SOSS eclipse spectrum of WASP-80\,b and the haze models from \cite{jacobs_probing_2023} with the respective $n_\sigma$ values computed from Equation \ref{eq:n_sigma} considering 27 degrees of freedom.}
\label{tab:hazes}
\begin{tabular}{|l||c|c||c|c||c|c||c|c|}
\hline
 & \multicolumn{4}{c||}{\textbf{Tholins}} & \multicolumn{4}{c|}{\textbf{Soots}} \\
\hline
Formation rate & \multicolumn{2}{c||}{[M/H]=0.0} & \multicolumn{2}{c||}{[M/H]=1.0} & \multicolumn{2}{c||}{[M/H]=0.0} & \multicolumn{2}{c|}{[M/H]=1.0} \\
\cline{2-9}
(g\,cm$^{-2}$\,s$^{-1}$) & $\chi^2_\nu$ & $n_\sigma$ & $\chi^2_\nu$ & $n_\sigma$ & $\chi^2_\nu$ & $n_\sigma$ & $\chi^2_\nu$ & $n_\sigma$ \\
\hline
\hline
1e-14 & 1.32 & 1.18 & 1.12 & 0.43 & 1.32 & 1.17 & 1.12 & 0.43 \\
\hline
3e-13 & 1.28 & 1.05 & 1.13 & 0.48 & 1.25 & 0.93 & 1.22 & 0.80\\
\hline
3e-12 & 1.42 & 1.53 & 1.29 & 1.06 & 1.11 & 0.41 & 1.19 & 0.69\\
\hline
3e-11 & 1.88 & 3.24 & 1.67 & 2.45 & 1.10 & 0.38 & 1.13 & 0.47\\
\hline
3e-10 & 1.92 & 3.39 & 1.72 & 2.65 & 1.12 & 0.44 & 1.14 & 0.50\\
\hline
\end{tabular}
\end{center}
\end{table}

\end{document}